\documentclass{aastex63}

\usepackage{amsmath}
\usepackage{times}
\usepackage{mathptmx}
\usepackage[caption=false]{subfig}
\DeclareMathAlphabet{\mathcal}{OMS}{cmsy}{m}{n}
\DeclareSymbolFont{largesymbols}{OMX}{cmex}{m}{n}

\newcommand{\LTE}{\emph{L-T-E} }
\newcommand{\LTEp}{\emph{L-T-E$_{\rm{p}}$} }

\newcommand{\Epi}{E_{\rm{p}}}
\newcommand{\Eiso}{E_{\gamma,\rm{iso}}}

\newcommand{\ext}{\sigma_{\rm{ext}}}

\newcommand{\dl}{d_{\rm{L}}}
\newcommand{\dH}{d_{\rm{H}}}
\newcommand{\dc}{d_{\rm{C}}}
\newcommand{\om}{\Omega_{m}}
\newcommand{\oL}{\Omega_{\Lambda}}
\newcommand{\ok}{\Omega_{k}}

\received{\today}
\revised{\today}
\accepted{XXX}

\shorttitle{de-evolved L-T-E$_{\rm{p}}$ correlation}
\shortauthors{Xu et al.}

\begin{document}

\title{X-ray Plateaus in Gamma-Ray Burst Afterglows and Their Application in Cosmology}

\correspondingauthor{Yong-Feng Huang (hyf@nju.edu.cn)}

\author[0000-0001-7943-4685]{Fan Xu}
\affiliation{School of Astronomy and Space Science, Nanjing University,
	Nanjing 210023, People's Republic of China}

\author{Chen-Han Tang}
\affiliation{School of Astronomy and Space Science, Nanjing University,
	Nanjing 210023, People's Republic of China}

\author{Jin-Jun Geng}
\affiliation{School of Astronomy and Space Science, Nanjing University,
	Nanjing 210023, People's Republic of China}

\author{Fa-Yin Wang}
\affiliation{School of Astronomy and Space Science, Nanjing University,
	Nanjing 210023, People's Republic of China}
\affiliation{Key Laboratory of Modern Astronomy and Astrophysics (Nanjing University),
	Ministry of Education, People's Republic of China}

\author{Yu-Yang Wang}
\affiliation{School of Astronomy and Space Science, Nanjing University,
	Nanjing 210023, People's Republic of China}

\author{Abudushataer Kuerban}
\affiliation{School of Astronomy and Space Science, Nanjing University,
	Nanjing 210023, People's Republic of China}

\author{Yong-Feng Huang$^\star$}
\affiliation{School of Astronomy and Space Science, Nanjing University,
	Nanjing 210023, People's Republic of China}
\affiliation{Key Laboratory of Modern Astronomy and Astrophysics (Nanjing University),
	Ministry of Education, People's Republic of China}

\begin{abstract}

For gamma-ray bursts (GRBs) with a plateau phase in the X-ray afterglow,
a so-called \LTE correlation has been found which tightly connects the
isotropic energy of the prompt GRB ($\Eiso$) with the end time of the
X-ray plateau ($T_{a}$) and the corresponding X-ray luminosity at the
end time ($L_{X}$). Here we show that there is a clear redshift evolution
in the correlation. Furthermore, since the power-law indices of $L_{X}$
and $\Eiso$ in the correlation function are almost identical, the \LTE
correlation is insensitive to cosmological parameters and cannot be used
as a satisfactory standard candle. On the other hand, based on a sample including 121
long GRBs, we establish a new three parameter correlation that connects
$L_{X}$, $T_{a}$ and the spectral peak energy $\Epi$, i.e. the \LTEp correlation.
This correlation strongly supports the so-called
Combo-relation established by \cite{Izzo..2015}.
After correcting for the redshift evolution, we show that
the de-evolved \LTEp correlation can be used as a standard candle.
By using this correlation alone, we are able to constrain the cosmological
parameters as $\om=0.389^{+0.202}_{-0.141}$ ($1\sigma$) for the flat
$\Lambda$CDM model, or $\om=0.369^{+0.217}_{-0.191}$,
$w=-0.966^{+0.513}_{-0.678}$ ($1\sigma$) for the flat $w$CDM model.
Combining with other cosmological probes, more accurate constraints on
the cosmology models are presented.

\end{abstract}

\keywords{cosmology: dark energy --- gamma-ray burst: general --- stars: neutron}

\section{Introduction} \label{sec:intro}

Since the discovery of the accelerating expansion of the Universe by using SNe Ia as
standard candles \citep{Phillips..1993,Riess..1998,Perlmutter..1999}, many studies
have given similar results through independent observations, including those of
the cosmic microwave background (CMB) \citep{Spergel..2003,Planck..2014,
Planck..2016,Planck..2018} and the baryon  acoustic oscillations
(BAO) \citep{Eisenstein..2005,SDSS-III..2017}.  These studies have presented the
so-called Lambda cold dark matter ($\Lambda$CDM) model as a standard cosmological
model, and the accelerating expansion of the Universe leads to a matter component
called the dark energy for which we still have little clue of its essence.
Even though these methods mentioned above are successful in many aspects,
deficiencies still exist.  The mechanism of a standard SN Ia limits its maximum
luminosity.  Therefore, the observed upper limit redshift of SNe Ia is around 2.  Meanwhile,
the CMB temperature anisotropy only provides information of the early Universe
at a high redshift around 1089 \citep{Planck..2018}.  Hence, there exists a
redshift gap between SNe Ia and CMB.

Gamma-ray bursts (GRBs) are extremely high-energy events with the luminosity ranging
from $10^{47}$ to $10^{54}$ erg s$^{-1}$ \citep{Zhangbing..2011} and redshifts
up to $z=9.4$ \citep{Cucchiara..2011}.  Therefore GRBs can potentially fill the gap between
SNe Ia and CMB.  In addition, comparing with the observations of SNe Ia which
suffer from the extinction of interstellar medium (ISM), gamma-ray photons are much
less affected during their propagation toward us \citep{Wang..2015}.  Thus,
many authors have tried to use GRBs as a complementary
probe \citep{Dai..2004,Liang..2008,Lin..2016,Wang..2016,Wang..2019,Amati..2019}.
All these studies are based on some empirical luminosity correlations
which can standardize GRBs \citep{Norris..2000, Fenimore..2000, Amati..2002,
Yonetoku..2004, Ghirlanda..2004,Liang..2005,Firmani..2006,Dainotti..2008,
Xu..2012}.  Recently, some authors obtained different results when
using high-redshift probes like GRBs.  For example, \cite{Amati..2019} combined
the $\Epi-\Eiso$ correlation of GRBs with supernova JLA data set and
found $\om=0.397_{-0.039}^{+0.040}$ at the $2\sigma$ level for the
flat $\Lambda$CDM model (also see \cite{Demianski..2019}).  This result is
over $3\sigma$ tension with that of \cite{Planck..2018}.

Recently \cite{Tang..2019} used a sample of 174 \emph{Swift} GRBs with
a plateau phase in the X-ray afterglow to restudy the so-called \LTE
correlation.  This correlation connects the end time of the plateau phase
in the GRB rest frame ($T_a$) and the corresponding X-ray luminosity at
that moment ($L_X$)  with the isotropic $\gamma$-ray energy release during
the prompt burst phase ($\Eiso$), and was first established by \cite{Xu..2012}.
\cite{Tang..2019} confirmed that there is a clear correlation among these
three parameters, i.e., $L_X \propto T_a^{-1.01}\Eiso^{0.84}$.
In this study, we will explore whether the \LTE correlation can be used
to probe the Universe. Additionally, considering that there is a tight
correlation between $\Eiso$ and the spectral peak energy $\Epi$ \citep{Amati..2002},
we will try to establish a new three parameter correlation involving $L_X$, $T_a$,
and $\Epi$, called the \LTEp correlation.  We will also explore the possibility
of using the \LTEp correlation to parameterizing the Universe.

When using empirical luminosity correlation of GRBs to constrain cosmological
parameters, one may confront the so-called ``circularity problem''.
Due to the paucity of local GRBs, one needs to assume a prior cosmological model to compute the luminosity distance, so that cosmological parameters derived directly from GRB correlations turn out to be model dependent. Several methods have been put forward to overcome this problem (see \cite{Dainotti..2017a,Dainotti..2018,Dainotti..2018b,Dainotti..book} for related reviews). \cite{Amati..2008} used a simultaneous fitting method, which can safely avoid this problem (also see \cite{Dainotti..2013a}). Other authors applied different calibration methods to find intrinsic GRB correlations. \cite{Ghirlanda..2006} used GRBs with redshifts in a narrow range to calibrate the $E_{\rm{p}}-E_{\gamma}$ (beaming-corrected $\gamma$-ray energy) relation, while \cite{Dainotti..2013} performed the Efron \& Petrosian method \citep{Efron..1992} to demonstrate the intrinsic nature of the $L_{X}-T_{a}$ correlation (i.e., the so-called Dainotti relation, see \cite{Dainotti..2008,Dainotti..2010,Dainotti..2011}). Recently some studies have used the Observational Hubble Dataset (OHD) to calibrate the GRB correlations \citep{Amati..2019,Luongo..2020a,Luongo..2020b}.  For the circularity problem, a most commonly used method is to engage SNe Ia data at low redshifts to calibrate GRBs as done
by \cite{Liang..2008} (also see: \cite{Schaefer..2007,Cardone..2009,Cardone..2010,Postnikov..2014}).
This method is based on the fact that objects at the
same redshift should have the same luminosity distance in any cosmology models.
When using SNe Ia to calibrate GRB luminosity correlation, we need to make sure
no redshift evolution exists in this correlation.  However, many other studies
have shown there could be a redshift evolution in some GRB empirical luminosity
correlations.  For example, \cite{Lin..2016} found that there could be redshift
evolution in several correlations (including the $\Epi-\Eiso$ correlation), but
no significant redshift evolution was found in $\Epi-E_{\gamma}$
relation. \cite{Demianski..2017} argued that they
found no redshift evolution in the $\Epi-\Eiso$ correlation (i.e., the so
called Amati relation, see \cite{Amati..2008}) with 162 GRB samples.
But in \cite{Wang..2017}, the Amati relation was found to evolve with redshift. As a result,
whether the Amati relation is redshift-dependent or not is still under debate.
Also, note that \cite{Dainotti..2013} have examined the redshift evolution in
the Dainotti relation and have overcome this problem with the application of reliable statistical methods.
Here, to compensate for the redshift evolution,
we will perform a calibration method to get a de-evolved \LTEp correlation.

It is worth noticing that both the \LTE and the \LTEp correlation are extensions
of the Dainotti relation.  Another example of the extension of the Dainotti relation is
the so-called fundamental plane relation, i.e., the $L_{X}-T_{a}-L_{\rm{peak}}$
correlation \citep{Dainotti..2016}.  This relation links the X-ray luminosity and the
duration of the plateau phase with the peak prompt luminosity. Very tight relationships
between these three parameters were found using class-specific GRB
samples \citep{Dainotti..2016,Dainotti..2017b,Dainotti..2020}.  In \cite{Dainotti..2020},
the intrinsic scatter of the fundamental plane relation with the platinum sample of
47 GRBs was found to be $\sigma_{\rm{platinum,cor}} = 0.22\pm0.10$.  This intrinsic
scatter is 44\% smaller than that of the \LTE correlation provided by \cite{Tang..2019}
using 174 GRBs.  Generally, the above correlations related to the plateau emission component
support the energy injection model in which a luminous millisecond magnetar is involved
to act as the central
engine \citep{Dall'Osso..2011,Xu..2012,Rowlinson..2014,Rea..2015,Li..2018,Stratta..2018}.

Meanwhile, it is interesting to note that a so-called Combo-relation was recently
established.  It is a four-parameter equation that involves $\alpha$, $L_X$, $T_a$, and $E_p$,
where $\alpha$ is the power-law timing index of the afterglow just after the plateau phase.
The Combo-relation can be regarded as an extension of the 2D $L_{X}-T_{a}$ Dainotti relation
\citep{Dainotti..2008,Dainotti..2010,Dainotti..2011,Dainotti..2013a,Dainotti..2015,Dainotti..2017a}
 and the Amati relation \citep{Amati..2008}, since the key parameters involved in it all
appear in the latter two relations.  
The detailed expression of the Combo-relation can be derived by jointly considering
the Amati relation \citep{Amati..2008} and a three parameter relation presented by \cite{Bernardini..2012}. 
According to the Amati relation \citep{Amati..2008}, $E_{\rm p}$ and
$E_{\gamma,\rm{iso}}$ are closely connected. At the same time, \cite{Bernardini..2012}
found that $E_{\rm p}$ and $E_{\gamma,\rm{iso}}$ are connected
with the X-ray energy release ($E_{X,\rm{iso}}$) in the afterglow phase.
From these two correlations,
\cite{Izzo..2015} concluded that there should be a direct $E_{X,\rm{iso}} -
E_{\rm p}$ relation. For those GRBs with a plateau in the X-ray afterglow,
they further argued that the term of $E_{X,\rm{iso}}$ should be calculated
by integrating all the energy release during the plateau phase and the
subsequent power-law decaying phase (also see \cite{Ruffini..2014}). 
Note that a correlation between $\alpha$ and $L_X$ was 
reported, which may provide interesting constraints on the origin of the 
plateau phase \citep{Dainotti..2015,Vecchio..2016}.   
As a result of the above deduction, the Combo-relation is expressed as
$L_X \propto E_p^{0.84 \pm 0.08} (T_a/|1+\alpha|)^{-1} $ \citep{Izzo..2015,Muccino..2021}.
It has been shown that the Combo-relation can be effectively used to probe the
Universe \citep{Izzo..2015,Muccino..2021,Luongo..2020a,Luongo..2020b}.
The \LTEp correlation studied here is quite
similar to the Combo-relation, but there is still difference between
them. In this study, we will also compare our results with those derived from
the Combo-relation.

Our paper is organized as follows.  In Section \ref{sec:LTE}, we briefly
review the the work of \cite{Tang..2019} and re-examine the \LTE correlation.
In Section \ref{sec:SN Ia}, the basic method we employed to
estimate the cosmological parameters is introduced. The \LTE correlation is then
used to try to constrain the cosmological parameters. However, it would be seen that
the result is not satisfactory. In Section \ref{sec:LTEp}, we established a new
three parameter correlation, i.e. the \LTEp correlation. Redshift evolution of
the \LTEp correlation is then examined, and a de-evolved \LTEp correlation is
further derived.  Constraints on the cosmological parameters by using the
de-evolved \LTEp correlation is presented in Section \ref{sec:LTEp fit}.
Finally, discussion and conclusions are given in Section \ref{sec:concl}.

\section{The L-T-E correlation} \label{sec:LTE}

The \LTE correlation was first established by \cite{Xu..2012}. Recently, it was confirmed
by \cite{Tang..2019} with a large sample of 174 \emph{Swift} GRBs.  It is interesting to
note that \cite{Zhao..2019} also independently confirmed the correlation at a high
confidence level. Here we attempt to use this correlation and the sample of \cite{Tang..2019}
to constrain the cosmological parameters.  It is worth noting that this sample contains seven
short GRBs which also follow the same correlation. All GRBs in the sample have a
plateau phase in the X-ray afterglow, which was fitted with a smoothly broken power-law
function of \citep{Evans..2009, Li..2012, Yi..2016}

\begin{equation}
	F_{X}(t)=F_{X0}\left[\left(\frac{t}{T_{0}}\right)^{\alpha_{1}\omega} +
	\left(\frac{t}{T_{0}}\right)^{\alpha_{2}\omega}\right]^{-1/\omega},
\end{equation}
where $T_{0}$ is the observed end time of the plateau phase that is related to $T_{a}$
as $T_{a}=T_{0}/(1+z)$;  $\alpha_{1}$, $\alpha_{2}$ are power-law indices of the plateau
phase and the subsequent decaying segment, respectively;  $F_{X0}$ is the flux at the
break time, $\omega$ describes the sharpness of the break, and the combination of these
two parameters gives the corresponding flux at the end of the plateau
as $F_{X}(T_{0})=F_{X0}\times2^{-1/\omega}$.

Consequently, the observed X-ray
luminosity at the end time of the plateau phase can be naturally derived as

\begin{equation} \label{eq:2}
	L_{X}'=4\pi \dl^{2}F_{X0}2^{-1/\omega},
\end{equation}
where $\dl$ is the luminosity distance,

\begin{equation} \label{eq:3}
	\dl = \left\{ \begin{array}{ll}
		\frac{\dH(1+z)}{\sqrt{-\ok}}\sin(\sqrt{-\ok}\frac{\dc}{\dH}), & \ok<0,\\
		(1+z)\dc, & \ok=0,\\
		\frac{\dH(1+z)}{\sqrt{\ok}}\sinh(\sqrt{\ok}\frac{\dc}{\dH}), & \ok>0,
	\end{array} \right.
\end{equation}
with $\dH=\frac{C_l}{H_{0}}$, $\ok=1-\om-\oL$, and $C_l$ is the speed of light.
The co-moving distance, $\dc$, is defined as

\begin{equation}
	\dc=\dH\int_{0}^{z} \frac{dz}{E(z)}=
\dH\int_{0}^{z} \frac{dz}{\sqrt{(1+z)^{3}\om+(1+z)^{3(1+w)}\oL+(1+z)^{2}\ok}}.
\end{equation}
In this section, $H_{0}=70.0$ km s$^{-1}$ Mpc$^{-1}$ along with $\om=1-\oL=0.286$ and $w=-1$
will be used to calculate the luminosity distance.

The so-called k-correction should be taken into account, i.e.,  $L_{X}=kL_{X}'$, where
$k$ is the coefficient of k-correction defined as

\begin{equation}
	k= \frac{\int_{e_{1}/(1+z)}^{e_{2}/(1+z)}E\phi(E)dE}{\int_{e_{1}}^{e_{2}}E\phi(E)dE}.
\end{equation}
Here $(e_{1},e_{2})$ brackets the energy band of the detector.
In our study, we use the same method as \cite{Tang..2019} by considering the spectrum as
a simple power-law function because of the relatively narrow energy range
possessed by BAT and XRT onboard \emph{Swift}.  Therefore, the corrected X-ray
luminosity should be

\begin{equation} \label{eq:6}
	L_{X}=\frac{4\pi \dl^{2}F_{X0}2^{-1/\omega}}{(1+z)^{2-\beta_{X}}},
\end{equation}
where $\beta_{X}$ is the photon spectral index.
The data of $\beta_{X}$ and $z$ can be found in the \emph{Swift} GRB
table\footnote{\url{https://swift.gsfc.nasa.gov/archive/grb\_table.html/}}.

Meanwhile, the isotropic energy of the prompt emission is

\begin{equation}
	\Eiso'=\frac{4\pi \dl^{2}S}{1+z},
\end{equation}
where $S$ is the BAT fluence.  Similarly, with the photon spectral index ($\alpha_{\gamma}$)
measured by BAT, the isotropic energy after considering the k-correction should be

\begin{equation} \label{eq:8}
	\Eiso=\frac{4\pi \dl^{2}S}{(1+z)^{3-\alpha_{\gamma}}}.
\end{equation}
The \LTE correlation can be written as
\begin{equation} \label{eq:9}
	\log{\frac{L_{X}}{10^{47}\rm{erg/s}}}=a + b \log{\frac{T_{a}}{10^{3}\rm{s}}}
      + c \log{\frac{\Eiso}{10^{53}\rm{erg}}}.
\end{equation}

\begin{figure}[htbp]
	\centering
	\subfloat{
		\includegraphics[width=0.8\linewidth]{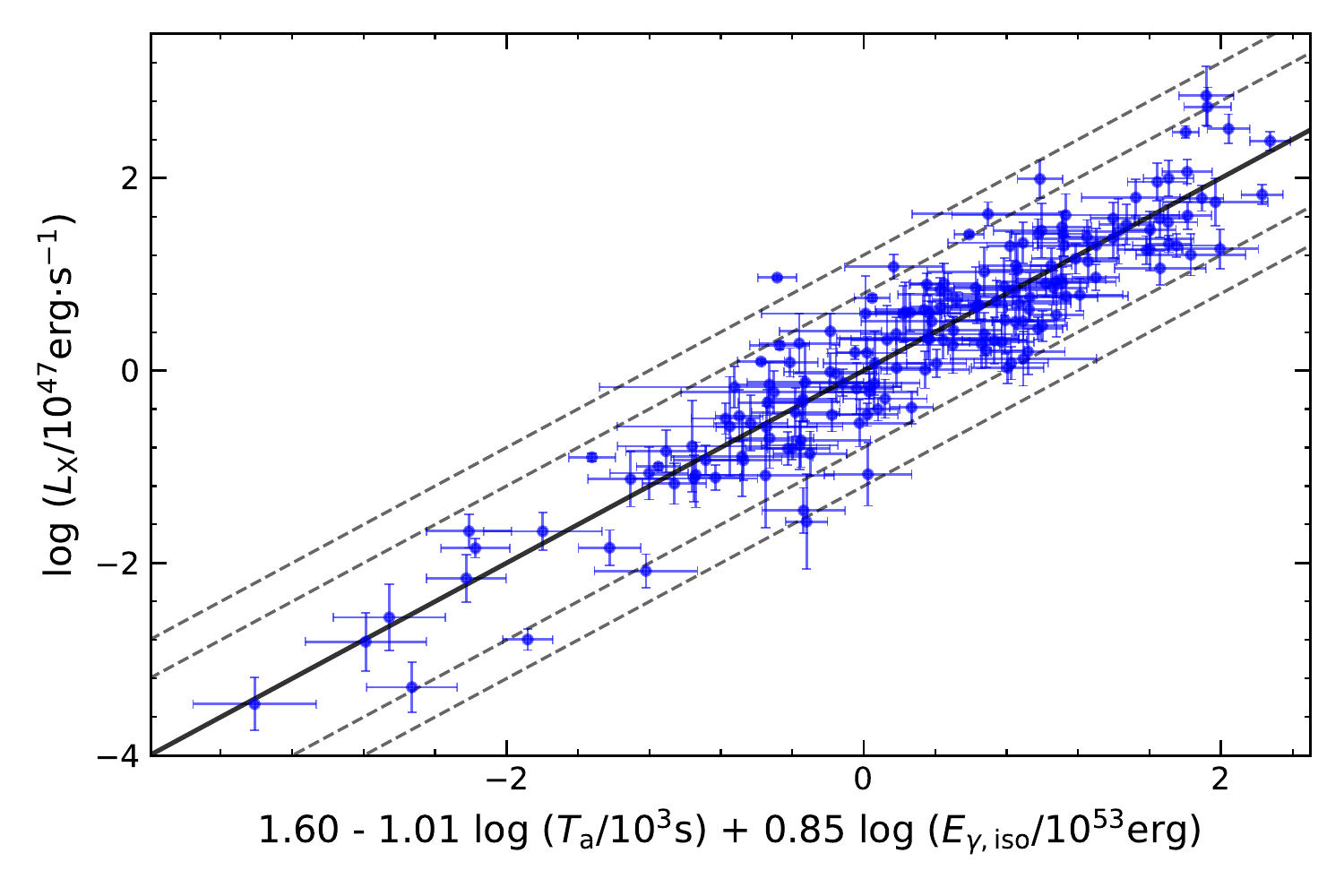}
	}
	
	\caption{The best-fit result of the \LTE correlation with the 174 GRBs taken
             from \cite{Tang..2019}. The solid line represents the best result
             and the dash lines represent $2\sigma$ and $3\sigma$
             confidence levels.}
	\label{fig:1}
\end{figure}

We use the Markov chain Monte Carlo (MCMC) algorithm to get the best-fit and
choose the likelihood implemented by \cite{2005} 
\begin{eqnarray} \label{eq:10}
	\lefteqn{ \mathcal{L}(a,b,c,\ext) \propto \prod_{i}
    \frac{1}{ \sqrt{ \ext^{2} + \sigma_{y_{i}}^{2} + b^{2}\sigma_{x_{1,i}}^{2}
            + c^{2}\sigma_{x_{2,i}}^{2} } }}
	\nonumber\\
	& & \times \exp \left[ -\frac{ \left( y_{i} - a - bx_{1,i} - cx_{2,i}
     \right)^{2} }{ 2\left( \ext^{2} + \sigma_{y_{i}}^{2}
     + b^{2}\sigma_{x_{1,i}}^{2} + c^{2}\sigma_{x_{2,i}}^{2} \right) }\right],
\end{eqnarray}
where $\ext$ is the extrinsic scatter parameter arouse from some hidden variables.
Note that \cite{Reichart..2001} proposed another form of the likelihood function, 
which is slightly different from D'Agostini's expression. 
Here we use D'Agostini's likelihood function in our calculations. 
We set $x_{1}=\log (T_{a}/10^{3} \rm{s})$, $x_{2}=\log (\Eiso/10^{53} \rm{erg})$
and $y=\log (L_{X}/10^{47} \rm{erg s}^{-1})$ and obtain the best-fit results
as $a=1.60\pm0.06$, $b=-1.01\pm0.05$, $c=0.85\pm0.04$ and $\ext=0.40\pm0.03$,
i.e., $L_X \propto T_a^{-1.01\pm0.05}\Eiso^{0.85\pm0.04}$.  The best-fit result
is illustrated in Figure \ref{fig:1}.
As a comparison, the \LTE correlation was originally derived by \citet{Xu..2012}
as $L_X \propto T_a^{-0.87\pm0.09}\Eiso^{0.88\pm0.08}$. It was then refined
as $L_X \propto T_a^{-1.01\pm0.05}\Eiso^{0.84\pm0.04}$ in \cite{Tang..2019}.
Also, \cite{Zhao..2019} derived the correlation as
$L_X \propto T_a^{-0.97\pm0.07}\Eiso^{0.79\pm0.05}$.
Our result is generally consistent with these studies in $1\sigma$ confidence level.

\section{Standardizing GRBs using SNe Ia} \label{sec:SN Ia}

In Section 2, we derived the \LTE correlation by assuming fixed values
for some cosmology parameters such as $H_0$, $\om$, $\oL$, and $w$. However,
in order to use this GRB sample to study the Universe, we first need to derive a
\LTE correlation that is independent of the cosmological models.
Here we choose the same method as presented by \cite{Liang..2008}, i.e.,
calibrating GRBs with SNe Ia. There are two main steps in the process:
1) use an available sample of SNe Ia to calibrate the distance modulus ($\mu$)
of the low-redshift GRBs and get the best-fit coefficients of the \LTE correlation
with the calibrated low-redshift GRBs; 2) use the fitting results and \LTE
correlation to calculate the model-independent $\mu'$ of the high-redshift GRBs
and get some constraints on cosmological parameters.

We use the so-called ``Pantheon sample'' of SNe Ia \citep{Pantheon..2018} to calibrate low-redshift
GRBs. This sample includes 1048 SNe Ia, of which the maximum redshift is 2.26.
However, except for one event with a redshift of 2.26, all other SNe Ia have a
redshift less than 2.0.  As a result, the ``Pantheon sample'' actually is effective
in calibrating our GRBs only in the region of $z<2.0$. Considering this ingredient,
we divide our GRB sample into two sub-samples according to the distance,
a low-redshift sample with 85 GRBs ($z<2$) and a high-redshift sample with 89 GRBs ($z>2$).
The boundary is $z=2.0$.

\subsection{Redshift evolution of the L-T-E correlation} \label{subsec:LTE evolution}

In order to be a satisfactory standard candle, the correlation itself should not
evolve at different redshifts. In this case, we can then safely apply the derived correlation,
calibrated at low redshifts, to high redshifts to constrain cosmological parameters.
If the correlation does evolve at different redshifts, then it could still be used only
when the evolution property is thoroughly known so that its effect can be compensated.
Otherwise, it cannot be used as a standard candle.

To test whether the \LTE correlation evolves with redshift, we can re-fit the three
parameter equation by using the two sub-groups, i.e. the low-redshift one and the
high-redshift one. To check it in more details, we have further grouped the GRBs
into four redshift bins, i.e., $z<1, 1<z<2, 2<z<3, 3<z$.  There are about 40 GRBs
in each bin. For these various sub-samples, the best-fit \LTE correlations are presented
in Table \ref{tab:1}.

\begin{table}[h!]
	\renewcommand{\thetable}{\arabic{table}}
	\centering
	\caption{Redshift evolution test of the $L_{x}-T_{a}-E_{\gamma,\rm{iso}}$ correlation.}
	\label{tab:1}
	\begin{tabular}{cccccc}
		\tablewidth{0pt}
		\hline
		\hline
		Redshift range & $a$     & $b$     & $c$     & $\ext$  & GRB number \\
		\hline
		Full data & 1.60 $\pm$ 0.06 & -1.01 $\pm$ 0.05 & 0.85 $\pm$ 0.04 & 0.40 $\pm$ 0.03 & 174 \\
		\hline
		$z<2$   & 1.45 $\pm$ 0.11 & -0.94 $\pm$ 0.08 & 0.84 $\pm$ 0.06 & 0.43 $\pm$ 0.04 & 85 \\
		$z>2$   & 1.51 $\pm$ 0.08 & -0.98 $\pm$ 0.07 & 0.56 $\pm$ 0.09 & 0.31 $\pm$ 0.04 & 89 \\
		\hline
		$z<1$   & 1.28 $\pm$ 0.20 & -0.97 $\pm$ 0.13 & 0.80 $\pm$ 0.09 & 0.52 $\pm$ 0.07 & 44 \\
		$1<z<2$ & 1.32 $\pm$ 0.14 & -0.84 $\pm$ 0.09 & 0.65 $\pm$ 0.11 & 0.31 $\pm$ 0.05 & 41 \\
		$2<z<3$ & 1.63 $\pm$ 0.12 & -1.07 $\pm$ 0.12 & 0.70 $\pm$ 0.14 & 0.29 $\pm$ 0.05 & 42 \\
		$3<z$   & 1.37 $\pm$ 0.10 & -0.84 $\pm$ 0.10 & 0.34 $\pm$ 0.13 & 0.32 $\pm$ 0.05 & 47 \\
		\hline
	\end{tabular}
\end{table}

\begin{figure}[htbp]
	\centering
	\subfloat{
		\includegraphics[width=0.8\linewidth]{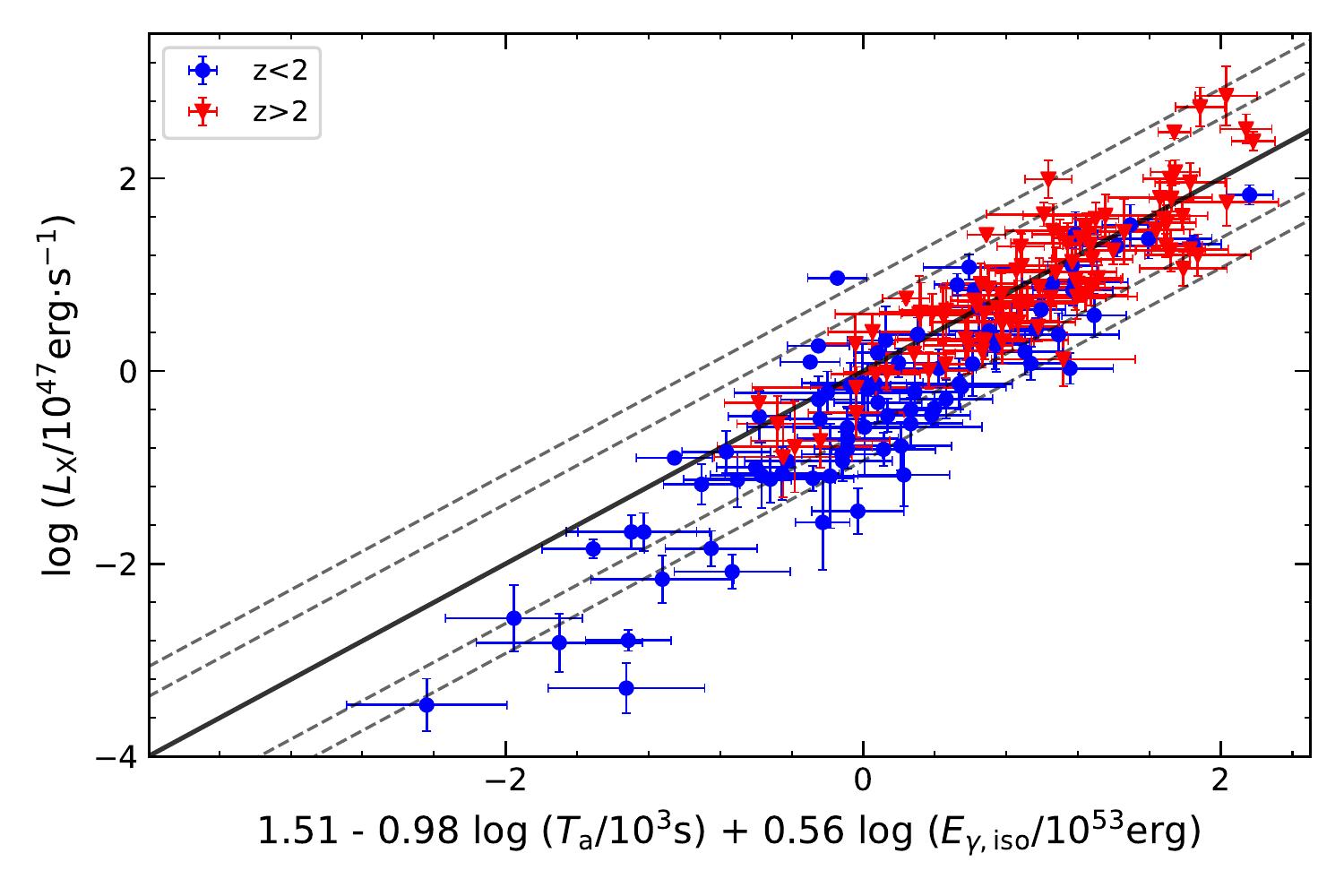}
	}
	\caption{The best-fit \LTE correlation of the high-redshift ($z>2$) GRB sub-sample
    (triangle points).  As a comparison, the low-redshift ($z<2$) GRBs are also plotted
    (dotted points).  The high-redshift GRBs distribute normally along the best-fit
    line, while the low-redshift events systematically deviate from the line.
    The dashed lines represent $2\sigma$ and $3\sigma$ confidence levels.}
	\label{fig:2}
\end{figure}

\begin{figure}[htbp]
	\centering
	\subfloat{
		\includegraphics[width=0.8\linewidth]{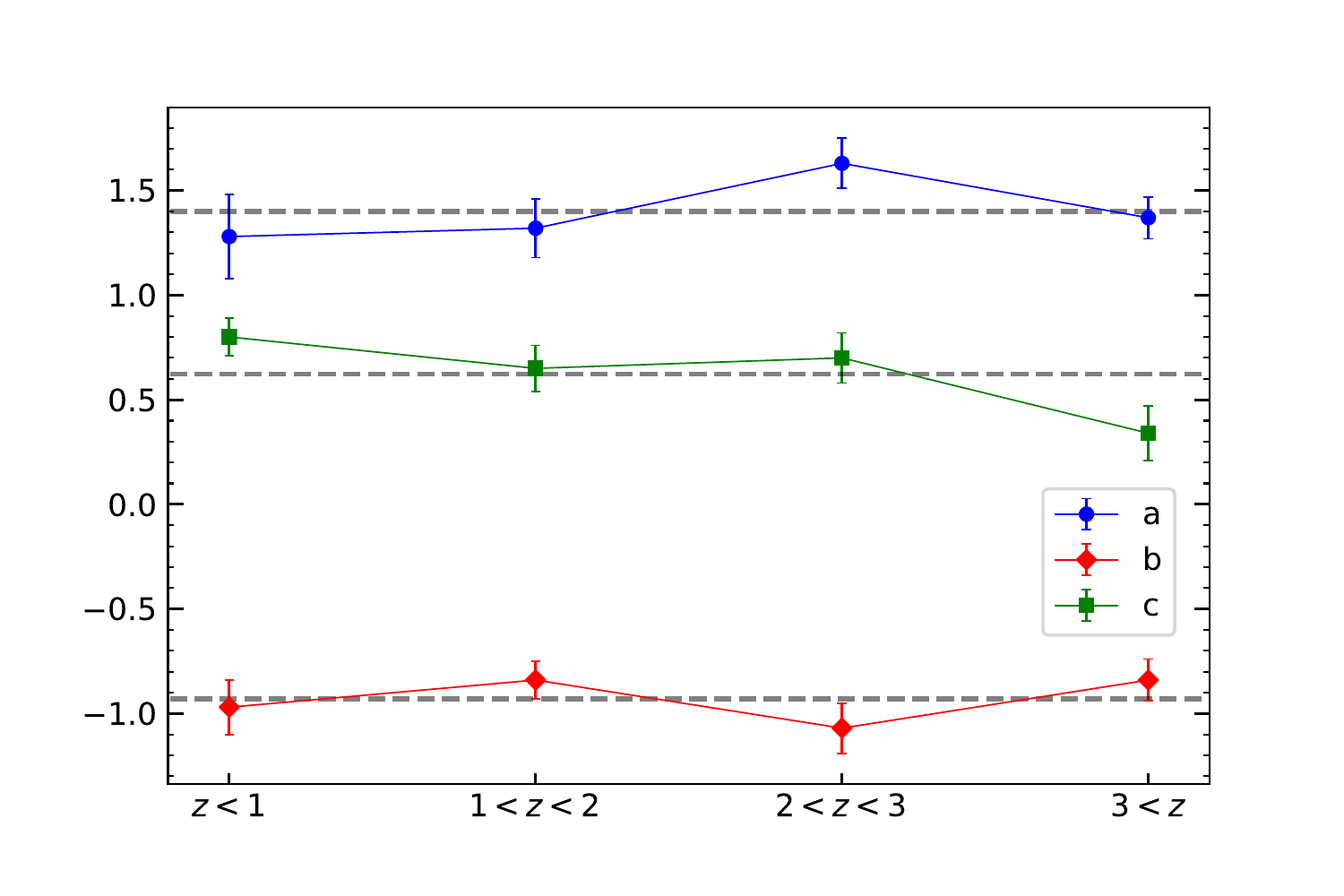}
	}
	
	\caption{The variation of the best-fit coefficients among different redshift bins.
    The dash lines represent the average values of the best-fit coefficients.}
	\label{fig:3}
\end{figure}

Table \ref{tab:1} indicates that the \LTE correlation indeed evolves with redshift. This
can be most clearly seen from the coefficient ``$c$'', which links the X-ray luminosity
and the isotropic energy. For the $z<2$ sub-sample, $c = 0.84 \pm 0.06$, but it changes to
$0.56 \pm 0.09$ when $z>2$. As for $a$ and $b$, their variation can almost be ignored
within the error bars. Anyway, the variation of a single coefficient indicates that
the whole \LTE correlation evolves with redshift. To make it more intuitive, using
the best-fit \LTE equation derived from the high-redshift sub-sample, we plot all
the low-redshift and high-redshift GRBs in Figure \ref{fig:2}. We see that while the
high-redshift GRBs distribute normally along the best-fit line, there is a systematic
deviation for the low-redshift events. When the GRBs are grouped into four redshift
bins, the evolution of the best-fit \LTE correlation can also be clearly seen. The
coefficient $c$ obviously decreases with the increase of the redshift. It varies from
$0.80 \pm 0.09$ for $z < 1$ to $0.34 \pm 0.13$ for $z > 3$.
In Figure \ref{fig:3}, we have plotted the best-fit coefficients of $a, b, c$
versus the redshift bins, which shows the decreasing tendency of $c$ directly.

\subsection{Cosmology from the L-T-E correlation} \label{subsec:LTE fit}

The redshift evolution indicates that the \LTE correlation cannot
act as a satisfactory standard candle. One may not expect encouraging
results on the cosmological parameters by using this correlation.
Anyway, assuming that the \LTE correlation could be used, we continue our
calculations to see how the results would be.

As the first step, we need to standardize GRBs and obtain a credible \LTE
correlation equation. We use a linear interpolation method to get
the distance modulus ($\mu$) of the low-redshift GRBs, i.e.,

\begin{equation}
	\mu_{\rm{GRB}}= \frac{z_{i+1} - z}{z_{i+1} - z_{i}} \mu_{i} +
	 \frac{z - z_{i}}{z_{i+1} - z_{i}} \mu_{i+1},
\end{equation}
where $\mu_{\rm{GRB}}$ represents the derived distance modulus of
a low-redshift GRB at $z$, while $\mu_{i+1}$ and $\mu_{i}$ are the distance
modulus of observed SNe Ia at the nearby redshifts $z_{i+1}$ and $z_{i}$,
respectively.  Using the corresponding error bars of nearby SNe Ia,
namely $\epsilon_{\mu,i+1}$ and $\epsilon_{\mu,i}$, the error bar of
the GRB distance modulus ($\sigma_{\mu}$) can be calculated from

\begin{equation}
	\sigma_{\mu}^{2}=\left( \frac{z_{i+1} - z}{z_{i+1} - z_{i}} \right)^{2} \epsilon_{\mu,i}^{2} +
	\left( \frac{z - z_{i}}{z_{i+1} - z_{i}} \right)^{2} \epsilon_{\mu,i+1}^{2}.
\end{equation}

After obtaining the distance modulus, we can get the luminosity distance from

\begin{equation} \label{eq:13}
	\mu= 5\log\frac{\dl}{\rm{Mpc}} + 25= 5\log\frac{\dl}{\rm{cm}} -97.45.
\end{equation}
Using this calibrated luminosity distance, we then go further to re-calculate
$L_{X}$ and $\Eiso$ of the low-redshift GRBs by using Equations~(\ref{eq:6}) and
(\ref{eq:8}). At this stage, we have the three key parameters
($L_{X}, T_{a}$, and $\Eiso$) of all low-redshift GRBs at hand and are ready to
re-derive a calibrated \LTE correlation. Again, we use the MCMC method to get
the best-fit result, which gives the coefficients as $a=1.41\pm0.11$, $b=-0.93\pm0.08$,
$c=0.83\pm0.06$, with an extrinsic scatter of $\ext=0.41\pm0.04$.
This calibrated correlation is independent of any cosmological models
and should be more reliable.

By directly extrapolating the calibrated \LTE correlation
to high-redshift GRBs, we can get their model-independent $\mu'$ through
Equations (\ref{eq:6}),(\ref{eq:8}),(\ref{eq:9}), and (\ref{eq:13}), i.e.,

\begin{equation} \label{eq:14}
	\mu'= \frac{5}{2(1-c)} \left[ a +b\log T_{a}
  +c\log\frac{4\pi S}{(1+z)^{3-\alpha_{\gamma}}}
  -\log\frac{4\pi F_{X0}2^{-1/\omega}}{(1+z)^{2-\beta_{X}}} \right] -97.45,
\end{equation}
where $S$ is in units of erg cm$^{-2}$ and $F_{X0}$ is in units of erg cm$^{-2}$ s$^{-1}$.
Correspondingly, the error bar of $\mu'$ can also be calculated by taking
the extrinsic scatter $\ext$ into account,
\begin{eqnarray}
	\lefteqn{ \sigma_{\mu'} = \frac{5}{2(1-c)} \Bigg\{ \ext^2
    +\sigma_{a}^2 +\sigma_{b}^2\log^{2} T_{a} +b^{2}\sigma^{2}_{\log T_{a}}
    +(\frac{\sigma_{F_{X0}}}{\ln10 \ F_{X0}})^2 +\log^{2}2 \ (\frac{\sigma_{\omega}}{\omega^{2}})^2
    + \sigma_{\beta_{X}}^{2}\log^{2}(1+z) }
	\nonumber\\
	& & +(\frac{\sigma_{c}}{1-c})^2 \left[ a +b\log T_{a}
    -\log\frac{F_{X0}2^{-1/\omega}}{(1+z)^{2-\beta_{X}}} \right]^{2}
    +(\frac{\sigma_{c}}{1-c})^2\log^{2}\frac{S}{(1+z)^{3-\alpha_{\gamma}}}
    +c^{2}\left[ (\frac{\sigma_{S}}{\ln10 \ S})^2
    +\sigma_{\alpha_{\gamma}}^{2}\log^{2}(1+z) \right] \Bigg\}^{1/2}.
\end{eqnarray}

In order to get the best-fit cosmological parameters, we maximize the
likelihood function,  $\mathcal{L}_{\rm{GRB}}(\theta)$, which is
constructed as \citep{Amati..2019}

\begin{equation} \label{eq:16}
		\mathcal{L}_{\rm{GRB}}(\theta)=\prod_{i=1}^{\mathcal{N}_{\rm{GRB}}}
    \frac{1}{\sqrt{2\pi}\sigma(z_{i})} \exp\left[-\frac{1}{2}
    \left( \frac{\mu'(z_{i})-\mu_{\rm{th}}(z_{i},\theta)}{\sigma(z_{i})}
    \right)^{2} \right],
\end{equation}
where $\theta$ represents a set of cosmological parameters and $\mu_{\rm{th}}$
can be derived from Equations (\ref{eq:3}) and (\ref{eq:13}).
$\mathcal{N}_{\rm{GRB}}$ here is the number of the high-redshift GRBs.

\begin{figure}[htbp]
	\centering
	\subfloat{
		\includegraphics[width=0.8\linewidth]{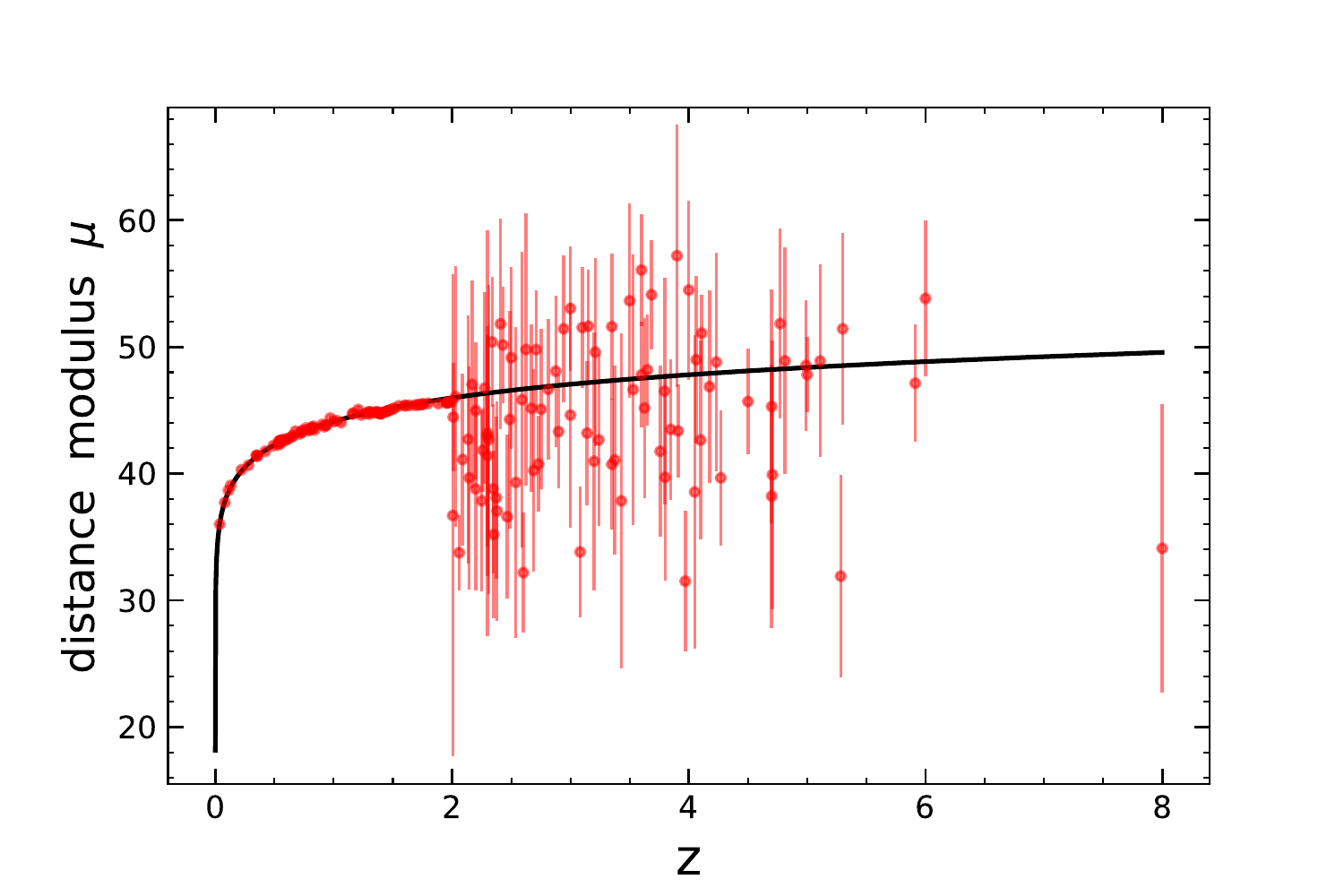}
	}
	
	\caption{Calibrated GRB Hubble diagram using the \LTE correlation.
    The data points represent our GRB sample. As a comparison, the solid curve
    corresponds to the theoretical distance modules calculated for a flat
    $\Lambda$CDM model with $H_{0}=70.0$ km s$^{-1}$ Mpc$^{-1}$ and $\om=0.289$. }
	\label{fig:4}
\end{figure}

Following the procedure described above, we have taken a test to see
whether the \LTE correlation can be used to constrain the cosmological
parameters or not. The Hubble diagram derived from our GRB sample is plotted
in Figure \ref{fig:4}. At low redshift of $z < 2$, we use the SNe Ia sample
to calibrate the \LTE correlation. The calibrated correlation is then
extended to high-redshift regime of $z > 2$. We see that the high-redshift
data points are highly dispersive with large error bars, which means they
can hardly give any meaningful constraints on the cosmological parameters.
There are at least two reasons for the fact that the \LTE correlation
fails to be a satisfactory cosmological probe. First, the \LTE correlation
is subjected to an obvious redshift evolution as described in the above
subsection. As a result, it is inappropriate to extrapolate the calibrated
low-redshift relation to high-redshift GRBs directly.
Second, and more importantly, both $L_{X}$ and $\Eiso$ are proportional
to the square of the luminosity distance, which itself is dependent on
cosmological parameters. At the same time, the best-fit value of the coefficient
$ c $ is close to 1.  It means the cosmological effect is largely canceled out
in the \LTE correlation, so that the correlation is insensitive to cosmological parameters.

\section{The L-T-E$_{\rm{p}}$ correlation} \label{sec:LTEp}

The \LTE correlation is not a good probe for cosmology. Here we explore the
possibility of finding another appropriate pattern. Noticing that the
spectral peak energy $\Epi$ is closely correlated with $\Eiso$ \citep{Amati..2002},
we conjecture that there may exist a correlation among $L_X$, $T_a$, and $\Epi$.
We write the potential relation as

\begin{equation} \label{eq:17}
	\log{\frac{L_{X}}{10^{47}\rm{erg/s}}}=a' + b' \log{\frac{T_{a}}{10^{3}\rm{s}}}
     + c' \log{\frac{\Epi}{\rm{keV}}},
\end{equation}
where $\Epi=E_{\rm{p,obs}}\times(1+z)$ and $E_{\rm{p,obs}}$ is the observed
peak energy in the $\nu F_{\nu}$ spectrum.
We now go on to derive the best-fit coefficients and assess the compactness of
this relation.
Note that the correlation among $L_X$, $T_a$, and $\Epi$ has been
explored by \cite{Izzo..2015}. In their studies, they have further included
a fourth parameter of $\alpha$ (the timing index of the power-law decaying
phase that follows the plateau segment) to get the so-called Combo-relation
as $L_X \propto (T_a/|1+\alpha|)^{-1} E_{\rm{p}}^{0.84 \pm 0.08}$ \citep{Izzo..2015,Muccino..2021}.
It can be regarded as a four-parameter relation.
In calibrating the relation, they have used a technique
that does not require a large sample of SNe Ia, thus it is less dependent
on other cosmological rulers \citep{Izzo..2015}. Additionally, the
Combo-relation is found to have no redshift evolution \citep{Muccino..2021}.
As a result, the relation has been effectively used to constrain cosmological
parameters \citep{Izzo..2015,Muccino..2021}.

Meanwhile, it is interesting to find that the fundamental plane relation,
i.e. the $L_{X}-T_{a}-L_{\rm{p}}$ correlation, is only one parameter different from
the \LTEp correlation.  This relation was first established by \cite{Dainotti..2016}
with a total sample of 176 Swift GRBs.  Also, they found the scatter of this relation
became smaller when using a class-specific GRB sample.  Recently, in \cite{Dainotti..2020},
they applied new criteria for GRB data sample, and obtained the so-called ``Platinum Sample''.
With this sample, they performed the Efron \& Petrosian method \citep{Efron..1992} to
remove the possible evolution in these three parameters and tried to find the intrinsic
relation between these three parameters.  Finally, a tight relationship between these
three parameters was derived
as $L_X \propto T_{a}^{-0.86 \pm 0.13} L_{\rm{p}}^{0.56 \pm 0.12}$ with a small
extrinsic scatter of about 0.22.

\subsection{Sample selection and the L-T-E$_{\rm{p}}$ data} \label{subsec:data}

Our new sample consist of the GRBs as investigated by \cite{Tang..2019}. We
select all the events with the $\Epi$ values available. For the parameter of
$\Epi$, we collect the data from the following sources:
\begin{itemize}
    \item Several papers involving the $\Epi - \Eiso$ correlation
          investigation \citep{Demianski..2017,Minaev..2020};
    \item The GRB catalogs of GBM/\emph{Fermi} \citep{Gruber..2014, Kienlin..2014,
          Bhat..2016,Kienlin..2020} and \emph{Swift} \citep{Lien..2016};
    \item The GCN circulars
          archive\footnote{\url{https://gcn.gsfc.nasa.gov/gcn3_archive.html}};
    \item A useful GRB database composed by \cite{Wang..2020}.
\end{itemize}
It should be noted that we omit all the short GRBs (7 events) included
in \cite{Tang..2019}, because the $\Epi - \Eiso$ correlation is different for
short and long GRBs \citep{Demianski..2017}.  Also, several GRBs have been
dropped out due to their poorly constrained $\Epi$. Finally we obtained a
sample with 121 long GRBs.  The details of this sample are listed
in Table \ref{tab:2}.

\subsection{L-T-E$_{\rm{p}}$ correlation} \label{subsec:LTEp evolution}

With the new data set, we now examine if there is a correlation
among $L_{X}$, $T_{a}$, and $\Epi$. We use the same MCMC method as described
in the above section. When all the events are engaged as a whole, we find that
the best-fit result is $a'=-1.03\pm0.37$, $b'=-1.08\pm0.08$, $c'=0.76\pm0.14$,
with an extrinsic scatter of $\ext=0.54\pm0.04$.
The fitting result is illustrated in Figure \ref{fig:5}. We see that there does
exist an obvious correlation among the three parameters, i.e.,
$L_{X} \propto T_a^{-1.08 \pm 0.08}\Epi^{0.76 \pm 0.14}$. We call this correlation
the \LTEp correlation.
Comparing with the Combo-relation of
$L_X \propto (T_a/|1+\alpha|)^{-1} E_p^{0.84 \pm 0.08}$ \citep{Izzo..2015, Muccino..2021},
we notice that the power-law indices of both $E_p$ and $T_a$
are interestingly consistent with each other in the two expressions.

Whether the new \LTEp correlation evolves with redshift should also be examined.
For this purpose, we divide the sample into two sub-samples,
the high-redshift sub-sample ($z>2$) and the low-redshift
sub-sample ($z<2$). We have re-fit the \LTEp correlation upon each sub-sample.
The results are shown in Table \ref{tab:3}. We see that the preferred value of $a$
is $-1.01$ when $z<2$, but it is 0.06 when $z>2$. Also, the preferred value of $c$
varies from 0.62 ($z<2$) to 0.44 ($z>2$). It indicates that the \LTEp correlation
also suffers from the redshift evolution.
This is somewhat unexpected, and is
different from the Combo-relation which is almost
redshift-independent \citep{Muccino..2021}.

\subsection{De-evolved L-T-E$_{\rm{p}}$ correlation} \label{subsec:de-evolved LTEp}

Three quantities ($L_{X}$, $T_{a}$, and $\Epi$) are involved in the \LTEp
correlation. Each quantity itself could be redshift
dependent \citep{Dainotti..2013} and the redshift evolution
of the whole relation thus might be a joint effect of the three quantities \citep{Demianski..2017}. To compensate for the redshift evolution, a natural idea is to assume that each quantity depends on
the redshift as a power-law function of $(1+z)$ with a certain
index \citep{Dainotti..2013}. However, since the \LTEp
correlation itself is a simple power-law function ($L_{X} \propto T_{a}^{b'}  \Epi^{c'}$),
we can also add a single power-law term of $(1+z)^{d'}$  into the correlation function
to synthesize the overall evolution effects of the three
quantities \citep{Demianski..2017}, i.e., the possible
redshift evolution of individual variable ($L_{X}$, $T_{a}$, and $\Epi$) is synthesized
in the new parameter of $d'$. Here we choose the simple power-law parameterization to simplify the calculation, different parameterization can be applied to flatten the function at higher redshifts \citep{Singal..2011,Dainotti..2015,Petrosian..2015}.
In this case, we re-write the \LTEp correlation as

\begin{equation} \label{eq:18}
	\log{\frac{L_{X}}{10^{47}\rm{erg/s}}}= a' +b'\log{\frac{T_{a}}{10^{3}\rm{s}}}
	+c'\log{\frac{\Epi}{\rm{keV}}} +d'\log(1+z).
\end{equation}
We use this new function to re-fit our data set. The result is also presented in
Table~\ref{tab:3}. It is found that the best-fit coefficient is $d'=1.78\pm0.29$,
which is significantly larger than 1. It indicates there really is a significant
redshift evolution in the direct \LTEp correlation. As a comparison, we have also
done a similar test for the \LTE correlation of Section \ref{sec:SN Ia}. The best-fit parameter
of $d'$ is only $d'=1.04 \pm 0.25$. It means that the redshift evolution of the \LTEp
correlation, if not corrected for, is even more significant than that of the previous
\LTE correlation.

\begin{figure}[htbp]
	\centering
	\subfloat{
		\includegraphics[width=0.5\linewidth]{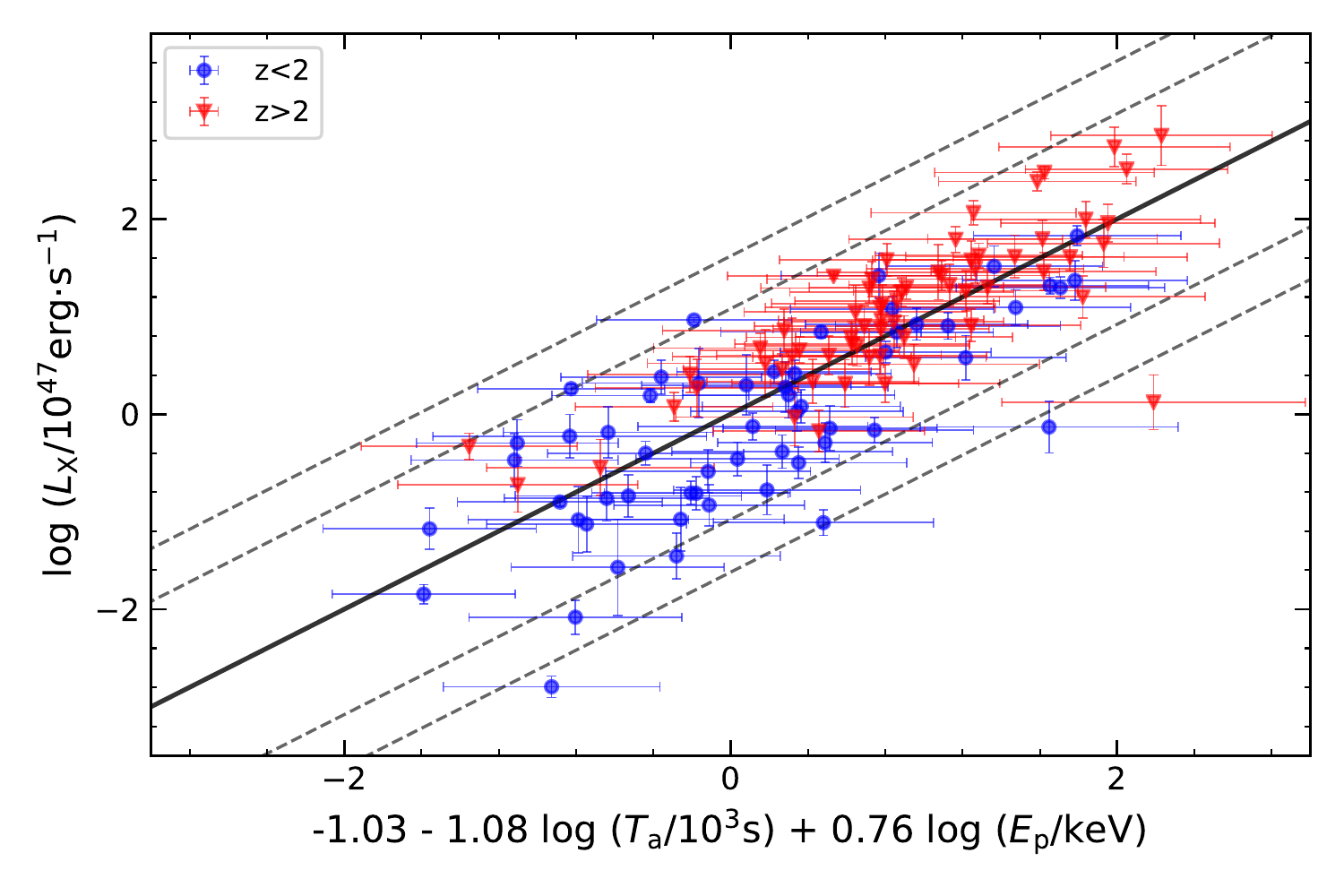}
	}
	\subfloat{
		\includegraphics[width=0.5\linewidth]{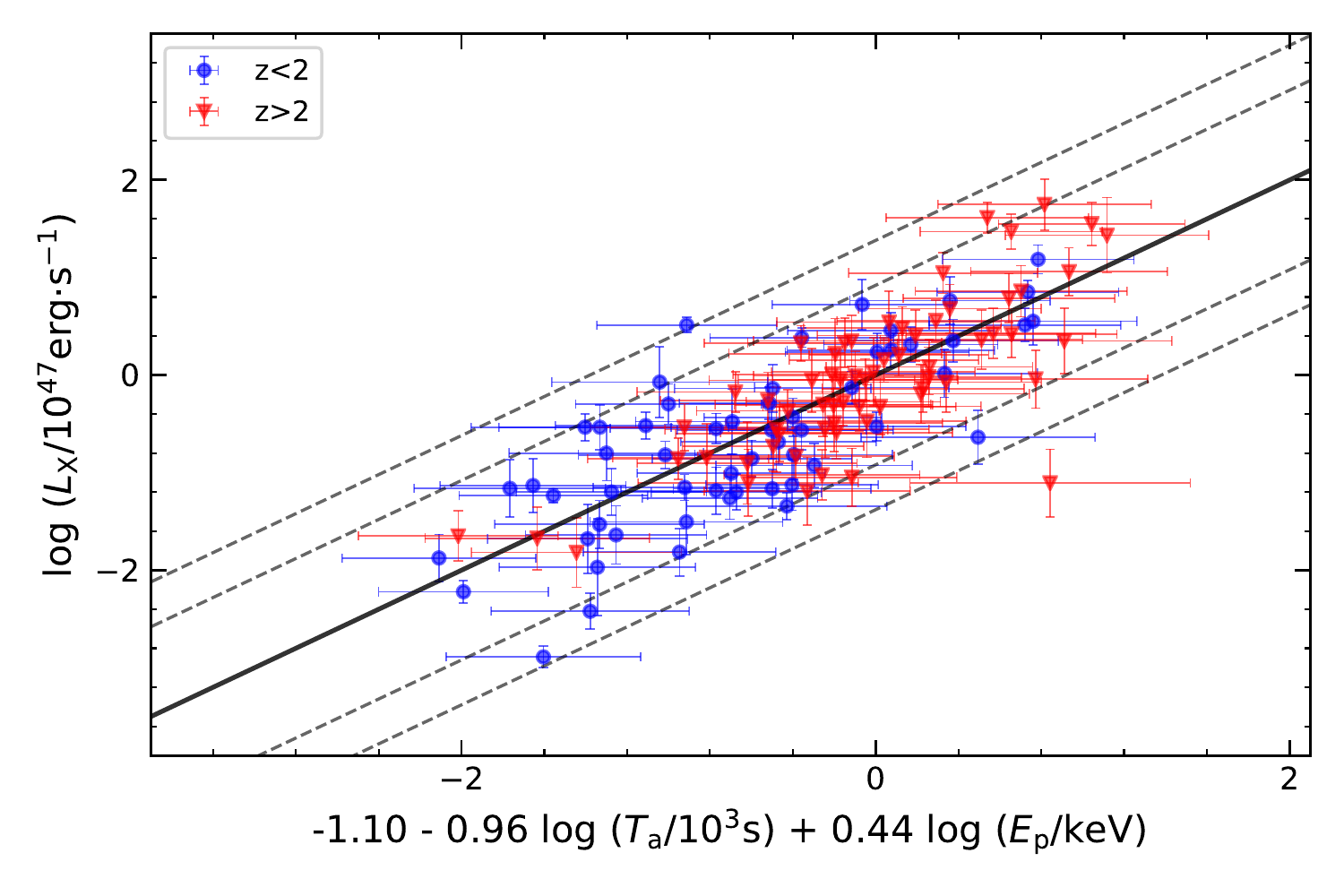}
	}
	\caption{The best-fit result of the \LTEp correlation (left panel)
   and the de-evolved \LTEp correlation (right panel). In the left panel, there
   is a clear systematic deviation between the low-redshift sub-sample and the
   high-redshift sub-sample. In the right panel, both sub-samples distributed
   normally along the best-fit line. }
	\label{fig:5}
\end{figure}

To rectify the evolution effect, we remove the redshift dependent term in
Equation \ref{eq:18} and replace $L_X$ with a calibrated X-ray luminosity of

\begin{equation} \label{eq:19}
	L_{X,\rm{cal}}=\frac{L_{X}}{(1+z)^{d'}}.
\end{equation}
The calibrated data of the X-ray luminosity are also listed in Table \ref{tab:2}
and the de-evolved \LTEp correlation can be written as

\begin{equation} \label{eq:20}
	\log{\frac{L_{X,\rm{cal}}}{10^{47}\rm{erg/s}}}= a'' +b''\log{\frac{T_{a}}{10^{3}\rm{s}}}
	+c''\log{\frac{\Epi}{\rm{keV}}}.
\end{equation}
Using this equation, the best-fit results for both low-redshift and high-redshift
GRBs are shown in Table \ref{tab:3}. Now, the derived coefficients are almost identical
for both sub-samples, which means no further redshift evolution exists.
In Figure \ref{fig:5}, we have plotted the best-fit results of the de-evolved \LTEp
correlation upon the full sample.  Comparing with the direct \LTEp correlation,
the de-evolved \LTEp correlation is more tighten and the data points of both the
low-redshift and high-redshift sub-samples are distributed normally along the
best-fit line.

We have further checked the possible evolution of each parameter.
For this purpose, we perform the Efron \& Petrosian method \citep{Efron..1992} to get the
redshift evolution of $L_{X}$, $T_{a}$, and $E_{\rm{p}}$ separately (details of this method are
presented in the Appendix). We take a simple power-law function $(1+z)^{k}$ as the evolution form
of each parameter.  Now we have three power-law indices, $k_{L_{X}}$, $k_{T_{a}}$, and $k_{E_{\rm{p}}}$
for $L_{X}$, $T_{a}$, and $E_{\rm{p}}$, respectively. Our analyses show that all the three
parameters evolve with redshift (see the Appendix for details). For $L_{X}$ and $T_{a}$, we
get $k_{L_{X}}=3.38\pm0.62$ and $k_{T_{a}}=-1.54\pm0.30$.  These two indices are somewhat different
from those derived by \cite{Dainotti..2017c}, in which both $L_{X}$ and $T_{a}$ have a smaller
evolution ($k_{L_{X}}=-0.40_{-0.83}^{+0.89}$ and $k_{T_{a}}=-0.17_{-0.37}^{+0.41}$).  The difference
may be due to the fact that different samples are used in the two studies.  Also, note that the error
bars derived in both studies are still quite large, further studies on this aspect are necessary when
more data are available in the future. As for $k_{E_{\rm{p}}}$, we get $k_{E_{\rm{p}}}=0.75\pm0.25$,
which also shows a significant redshift evolution.

After correcting for the redshift evolution of each parameter, we can also get a new separate de-evolved \LTEp correlation as

\begin{equation}
	\log{\frac{L_{X}/(1+z)^{k_{L_{X}}}}{10^{47}\rm{erg/s}}}= a''' +b'''\log{\frac{T_{a}/(1+z)^{k_{T_{a}}}}{10^{3}\rm{s}}}
	+c'''\log{\frac{E_{\rm{p}}/(1+z)^{k_{E_{\rm{p}}}}}{\rm{keV}}}.
\label{eq:21}
\end{equation}
Using the MCMC method, we get the best-fit coefficients for this correlation
as $a'''=-0.88\pm0.36$, $b'''=-0.87\pm0.10$, $c'''=0.35\pm0.16$, and $\sigma=0.44\pm0.05$.
The extrinsic scatter is smaller compared with the direct \LTE correlation, which indicates
that this new correlation does have made compensation for the redshift evolution to some
extent.

\section{Cosmology with de-evolved L-T-E$_{\rm{p}}$ correlation} \label{sec:LTEp fit}

In this section we use the de-evolved \LTEp correlation to constrain the cosmological
parameters.  It is well known that there is a problem called ``the Hubble tension'',
which means even the Hubble constant is not well determined.
For example, the Hubble constant derived from \cite{Planck..2018} was
$H_{0}=67.4\pm0.5$ km s$^{-1}$ Mpc$^{-1}$, while \cite{Riess..2019} used the
distance-ladder measurements of LMC Cepheids and got $H_{0}=74.03 \pm 1.42$
km s$^{-1}$ Mpc$^{-1}$.  In our study, we fix this parameter as $H_{0}=70.0$
km s$^{-1}$ Mpc$^{-1}$ for simplicity.  First, using the same method
as described in Section \ref{subsec:LTE fit}, we do a linear interpolation with the
help of the SNe Ia sample to get the calibrated $\mu$ of the low-redshift GRBs.
From the low-redshift GRBs, the best-fit result
of the de-evolved \LTEp correlation corresponds to $a''=-1.34\pm0.50$, $b''=-0.89\pm0.11$,
$c''=0.50\pm0.20$ and $\ext=0.52\pm0.07$.  Thus, the de-evolved \LTEp correlation can
be written as

\begin{equation}
	L_{X,\rm{cal}} \propto T_a^{-0.89\pm0.11}\Epi^{0.50\pm0.20}.
\label{eq:22}
\end{equation}
Because this de-evolved \LTEp correlation does not evolve with redshift,
it is applicable for the high-redshift sample.

\begin{figure}[htbp]
	\centering
	\includegraphics[width=0.8\linewidth]{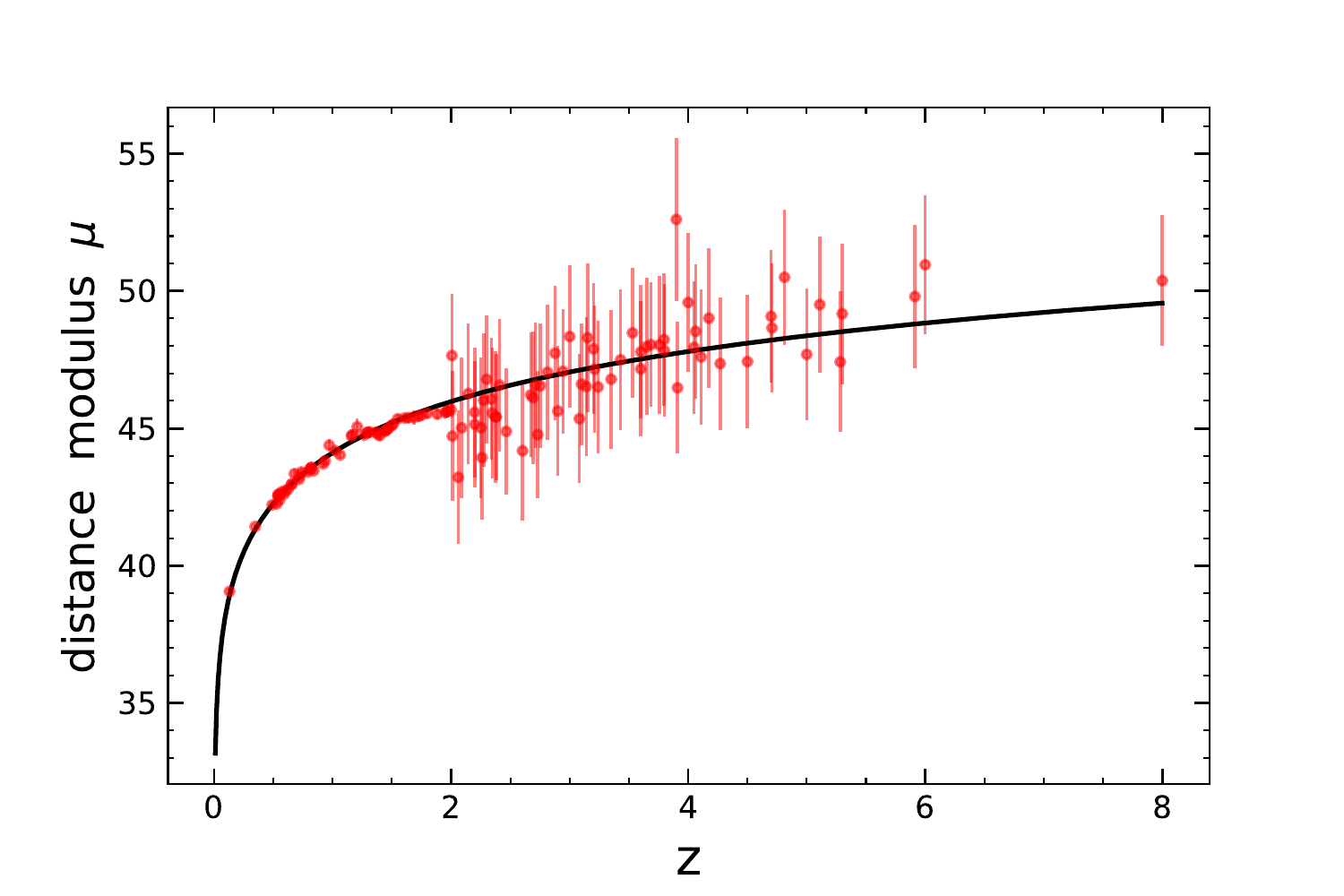}
	\caption{Calibrated GRB Hubble diagram from the de-evolved \LTEp correlation.  The data points
       correspond to our GRB sample.  As a comparison, the black line is the theoretical
       distance modulus calculated for the flat $\Lambda$CDM model
       with $H_{0}=70.0$ km s$^{-1}$ Mpc$^{-1}$ and $\om=0.289$.}
	\label{fig:6}
\end{figure}

The $L_{X}$ values of the high-redshift GRBs can be derived from Equations (\ref{eq:19})
and (\ref{eq:20}).   The propagated uncertainties of $L_{X}$ can be calculated from

\begin{equation}
	\sigma_{\log L_{X}}^{2}= \sigma_{a''}^{2} +\sigma_{b''}^{2}\log^{2}T_{a}
    +b''^{2}\sigma^{2}_{\log T_{a}} +\sigma_{c''}^{2}\log^{2}\Epi
    +c''^{2}\sigma^{2}_{\log \Epi} +\sigma_{d'}^{2}\log^{2}(1+z) +\ext^{2}.
\end{equation}
The model-independent $\mu'$ is calculated through Equations (\ref{eq:6}), (\ref{eq:13}),
and (\ref{eq:19}).  The final uncertainty of the distance modulus is given by

\begin{equation}
	\sigma_{\mu'}= \frac{5}{2}\left[ \sigma_{\log L_{X}}^{2}
    +(\frac{\sigma_{F_{X0}}}{\ln10 \ F_{X0}})^2 +\log^{2}2 \ (\frac{\sigma_{\omega}}{\omega^{2}})^2
    +\sigma_{\beta_{X}}^{2}\log^{2}(1+z) \right]^{1/2}.
\end{equation}
In Figure \ref{fig:6}, we present the Hubble diagram as built by using the calibrated
distance modulus.  With the model-independent $\mu'$ of the high-redshift GRBs, we can get
some constraints on the cosmological parameters by using Equation~(\ref{eq:16}). At the
same time, we should also note that the error bars of the derived distance modulus of the
high-redshift GRBs are still a bit too large, it means the cosmological parameters
cannot be accurately measured by simply using the devolved \LTEp correlation alone
currently. It can only be used as a supplementary tool.

Anyway, as the first step, we have made a test to see what constraints can be derived
by using only the GRBs. In this case, for the flat $\Lambda$CDM model, we obtain
$\om=0.389^{+0.202}_{-0.141}$.  When considering a non-flat $\Lambda$CDM model, the
best-fit result gives $\om=0.333^{+0.188}_{-0.142}$ and $\oL=0.346^{+0.356}_{-0.249}$.
Furthermore, if the flat $w$CDM model is considered, where $w$ describes the dark energy
equation of state (EOS), then we get $\om=0.369^{+0.217}_{-0.191}$
and $w=-0.966^{+0.513}_{-0.678}$. Note that the EOS parameter ($w$) is very close to the
standard value of $-1$. However, in all these results, the uncertainties of the derived
parameters are large.  The results are summarized in Table \ref{tab:4}.

As mentioned before, the circularity problem is a key issue that should
be paid special attention to when using various standard candles to measure the Universe.
For the de-evolved \LTEp correlation, an interesting method that can help overcome the difficulty
is to simultaneously derive the coefficients of the relation and the cosmological parameters
by fitting the observational data. In this case, no cosmological models would be assumed in
the first place and this simultaneous fitting method does not suffer from the circularity problem.
In our study, we have also tried this method to further check our results.

Following \cite{Dainotti..2013a}, the likelihood function will be similar to Equation (\ref{eq:10}) and we only need to add extra cosmological parameters into the parameter space.  
For the sake of a direct comparison, here we still use the D'Agostini's likelihood \citep{2005}, 
but not the Reichart's likelihood  \citep{Reichart..2001}.  The likelihood function now becomes

\begin{eqnarray}
	\lefteqn{ \mathcal{L}(a,b,c,\ext,p_{c}) \propto \prod_{i}
	\frac{1}{ \sqrt{ \ext^{2} + (\sigma_{L_{X}}^{i})^{2}
			+ b^{2}(\sigma_{T_{a}}^{i})^{2} + c^{2}(\sigma_{E_{\rm{p}}}^{i})^{2} } }}
	\nonumber\\
	& & \times \exp \left[ -\frac{ \left( L_{X}^{i}(p_{c}) - a - bT_{a}^{i} - cE_{\rm{p}}^{i}
	\right)^{2} }{ 2\left( \ext^{2} + (\sigma_{L_{X}}^{i})^{2}
	+ b^{2}(\sigma_{T_{a}}^{i})^{2} + c^{2}(\sigma_{E_{\rm{p}}}^{i})^{2} \right) }\right],
\end{eqnarray}
where $p_{c}$ refers to a set of cosmological parameters.  Here the cosmological parameters are free and need to be determined together with the correlation coefficients. 

\begin{figure}[htbp]
	\centering
	\includegraphics[width=0.8\linewidth]{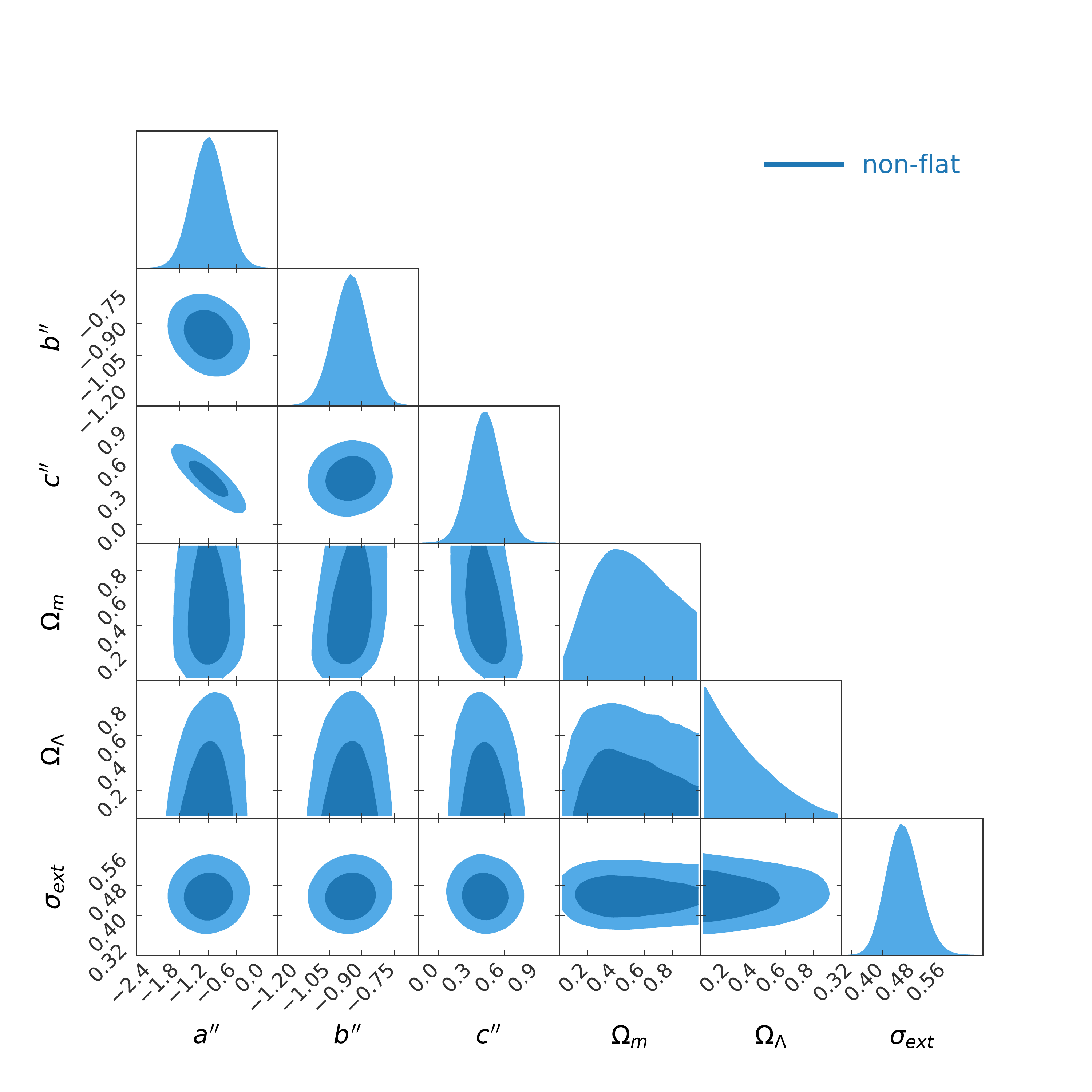}
	\caption{De-evolved \LTEp correlation coefficients and cosmological parameters derived
by using the simultaneous fitting method, which is not subjected to the so-called circularity
problem. This figure is plotted for a non-flat $\Lambda$CDM model (68\% and 95\% confidence levels
are shown).  The best-fit correlation coefficients are $a''=-1.19\pm0.33$, $b''=-0.95\pm0.07$,
$c''=0.43\pm0.13$, and $\ext=0.45\pm0.04$.  For the cosmological parameters, we
have $\om=0.50^{+0.30}_{-0.26}$, $\oL=0.23^{+0.29}_{-0.17}$. Here the error bars correspond to $1\sigma$ range. }
	\label{fig:7}
\end{figure}

\begin{figure}[htbp]
	\centering
	\includegraphics[width=0.8\linewidth]{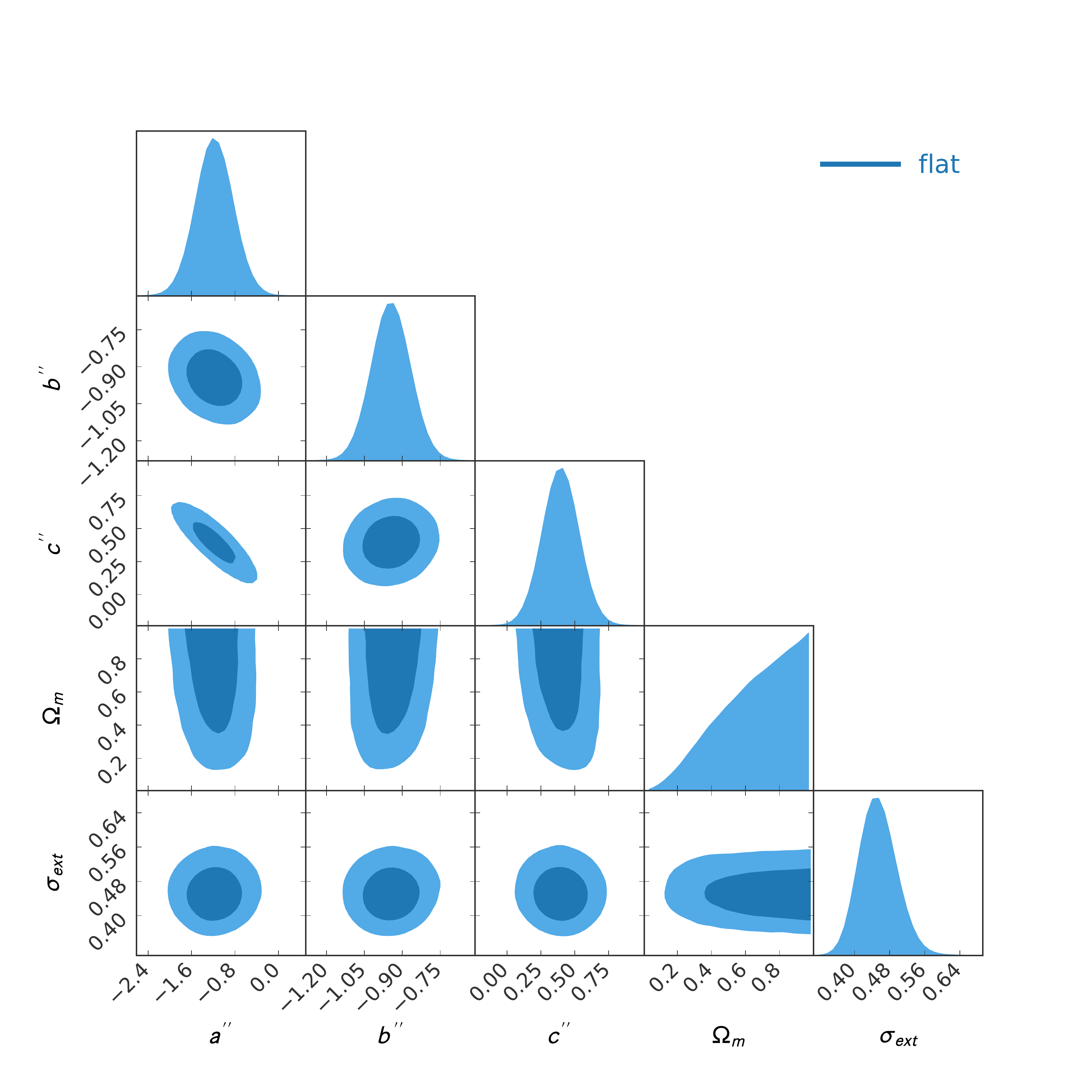}
	\caption{De-evolved \LTEp correlation coefficients and cosmological parameters derived
by simultaneously fitting the correlation coefficients and the cosmological parameters.  Note that this figure is different from Figure \ref{fig:7}. It is plotted for a flat $\Lambda$CDM model (68\% and 95\% confidence levels are shown).  The best-fit correlation coefficients are $a''=-1.18\pm0.32$, $b''=-0.95\pm0.07$,
$c''=0.40\pm0.13$, and $\ext=0.45\pm0.04$, while the best-fit result for the cosmological parameter is $\om=0.71^{+0.20}_{-0.28}$. Here the error bars correspond to $1\sigma$ range. }
	\label{fig:8}
\end{figure}

Using this likelihood function, we have re-fitted the observed plateau data.
For a non-flat $\Lambda$CDM model, our best-fit results are illustrated in Figure~\ref{fig:7}.
We find that the de-evolved \LTEp correlation now
becomes $L_{X,\rm{cal}} \propto T_a^{-0.95\pm0.07}\Epi^{0.43\pm0.13}$. This expression is
well consistent with the above result of Equation~(\ref{eq:22}). Correspondingly, the
derived cosmological parameters are $\om=0.50^{+0.30}_{-0.26}$ and $\oL=0.23^{+0.29}_{-0.17}$.
It is also consistent with the results presented above.
Similarly, for a flat universe, by using the simultaneous fitting method, we get the
cosmological parameters as $\om = 0.62^{+0.26}_{-0.35}$ and $w = -0.65^{+0.47}_{-0.86}$ for
the $w$CDM model, and $\om=0.71^{+0.20}_{-0.28}$ for the $\Lambda$CDM model.  We also plot the best-fit results for the flat $\Lambda$CDM model in Figure \ref{fig:8}.  These results
are all consistent with the above results within $1\sigma$ confidence
level, but with a larger uncertainty.  It clearly shows that our previous calibration and 
de-evolution processes are credible.  We will continue to use our previous calibration method in the following study for simplicity and clarity.

\begin{figure}[htbp]
	\centering
	\includegraphics[width=0.8\linewidth]{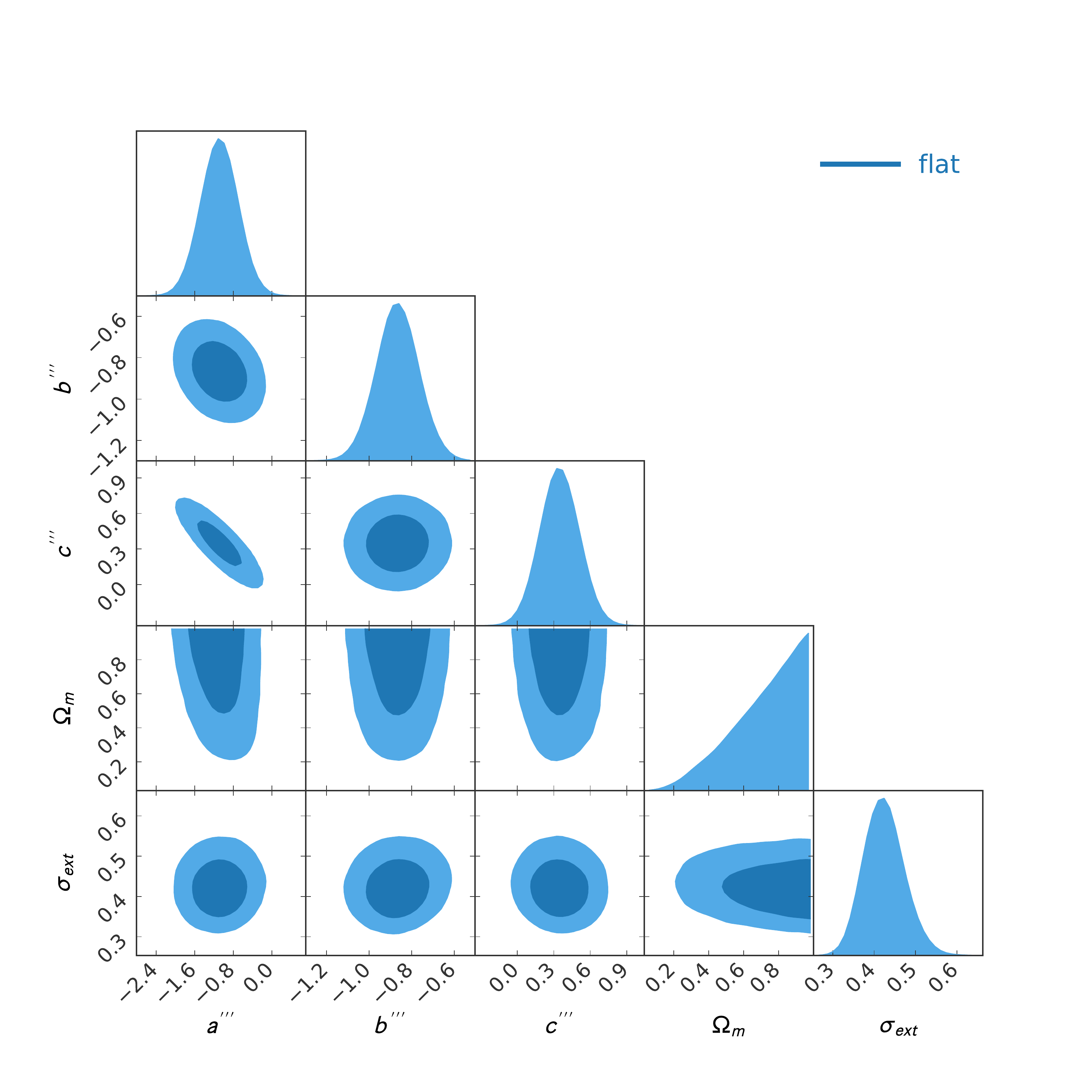}
	\caption{Best-fit correlation coefficients and cosmological parameters derived by using
the simultaneous fitting method and the separate de-evolved \LTEp correlation (Equation \ref{eq:21}).
A flat $\Lambda$CDM model is assumed. 68\% and 95\% confidence levels are shown in the plot.
The best-fit correlation coefficients are $a'''=-1.09\pm0.35$, $b'''=-0.86\pm0.10$, $c'''=0.35\pm0.15$,
and $\ext=0.42\pm0.04$. The cosmological parameter is derived as $\om=0.76_{-0.26}^{+0.17}$. }
	\label{fig:9}
\end{figure}

With the separate de-evolved \LTEp correlation (Equation \ref{eq:21}), we have also
attempted to use the simultaneous fitting method to constrain cosmological parameters.
For the flat $\Lambda$CDM model, the best-fit results are illustrated in Figure \ref{fig:9}.
We see that the matter density is derived as  $\om=0.76_{-0.26}^{+0.17}$. This result is
consistent with the former result using the de-evolved \LTEp correlation within $1\sigma$ confidence
level. Below, we will continue our study in our former framework
with the de-evolved \LTEp correlation (using SNe Ia calibration) for simplicity.

Next we combine the devolved \LTEp correlation with other cosmological
probes, including SNe Ia, CMB, and BAO, to give more stringent constraints
on the cosmological parameters. For this purpose, we should effectively
synthesize the likelihood functions of various probes.

The likelihood function of SNe Ia is calculated as

\begin{equation}
	\mathcal{L}_{\rm{SN}}(\theta)= \prod_{i=1}^{\mathcal{N}_{\rm{SN}}}
    \frac{1}{\sqrt{2\pi}\sigma_{\rm{SN}}(z_{i})} \exp\left[-\frac{1}{2}
    \left( \frac{\mu_{\rm{SN}}(z_{i})-\mu_{\rm{th}}(z_{i},\theta)}
    {\sigma_{\rm{SN}}(z_{i})} \right)^{2} \right],
\end{equation}
where $\mu_{\rm{SN}}$ and $\sigma_{\rm{SN}}(z_{i})$ are taken from the
Pantheon SNe Ia sample \citep{Pantheon..2018}.

For the CMB, we consider a shift parameter
$\mathcal{R}$, which is defined as \citep{Wang..2006}

\begin{equation}
	\mathcal{R}= \frac{\sqrt{\om}}{\sqrt{|\ok|}}
    \rm{sinn}\left( \sqrt{|\ok|}\int_{0}^{z_{\ast}} \frac{dz}{E(z)} \right),
\end{equation}
where $z_{\ast}$ is the last scattering redshift which we take as
$z_{\ast}=1089.90\pm0.23$ \citep{Planck..2016}. The function sinn($x$) is
defined as sinn($x$)=sin($x$) for a closed Universe, sinn($x$)=sinh($x$)
for an open Universe, and sinn($x$) = $x$ for a flat Universe.
The likelihood function for CMB is then

\begin{equation}
	\mathcal{L}_{\rm{CMB}}(\theta)=
    \frac{1}{\sqrt{2\pi}\sigma_{\mathcal{R}}} \exp\left[-\frac{1}{2}
    \left( \frac{\mathcal{R}(\theta)-\mathcal{R}_{\rm{obs}}}
    {\sigma_{\mathcal{R}}} \right)^{2} \right],
\end{equation}
where $\mathcal{R}_{\rm{obs}}$ and $\sigma_{\mathcal{R}}$ are constraints
obtained from the 2015 Planck data.
We take $\mathcal{R}_{\rm{obs}}=1.7474\pm0.0051$ for a non-flat Universe
and $\mathcal{R}_{\rm{obs}}=1.7482\pm0.0048$ for a flat Universe \citep{Wang.y..2016}.

As for the BAO, we use the data provided by \cite{SDSS-III..2017} and
the likelihood for BAO can be generally calculated
as \citep{matrix..2017,Tu..2019}

\begin{equation}
	\mathcal{L}_{\rm{BAO}}(\theta) \propto \frac{1}{\sqrt{2\pi}}
    \exp\left[-\frac{1}{2} (V_{\rm{obs}} -V_{\rm{th}})C^{-1}(V_{\rm{obs}}
    -V_{\rm{th}})^{T} \right],
\end{equation}
where the subscripts ``obs'' and ``th'' represent observed value and
theoretical value respectively. Here $V$ can be either the transverse
co-moving distance $D_{M}$ or the Hubble parameter $H_{z}$.  $C^{-1}$ is the inverse covariance
matrix of the observed variables.  We list the covariance matrix and the
observational data in Table \ref{tab:5} \citep{SDSS-III..2017}.
Note that $D_{M}$ in Table~\ref{tab:5} can be calculated
as $D_{M}=\dl/(1+z)$, and $H_{z}$ is defined as $H_{z}=H_{0}E(z)$.

To combine all the probes described above, we can define a joint likelihood
function as $\mathcal{L}=\mathcal{L}_{\rm{GRB}}
\mathcal{L}_{\rm{SN}} \mathcal{L}_{\rm{CMB}} \mathcal{L}_{\rm{BAO}}$.


\begin{figure}[htbp]
	\centering
	\includegraphics[width=0.8\linewidth]{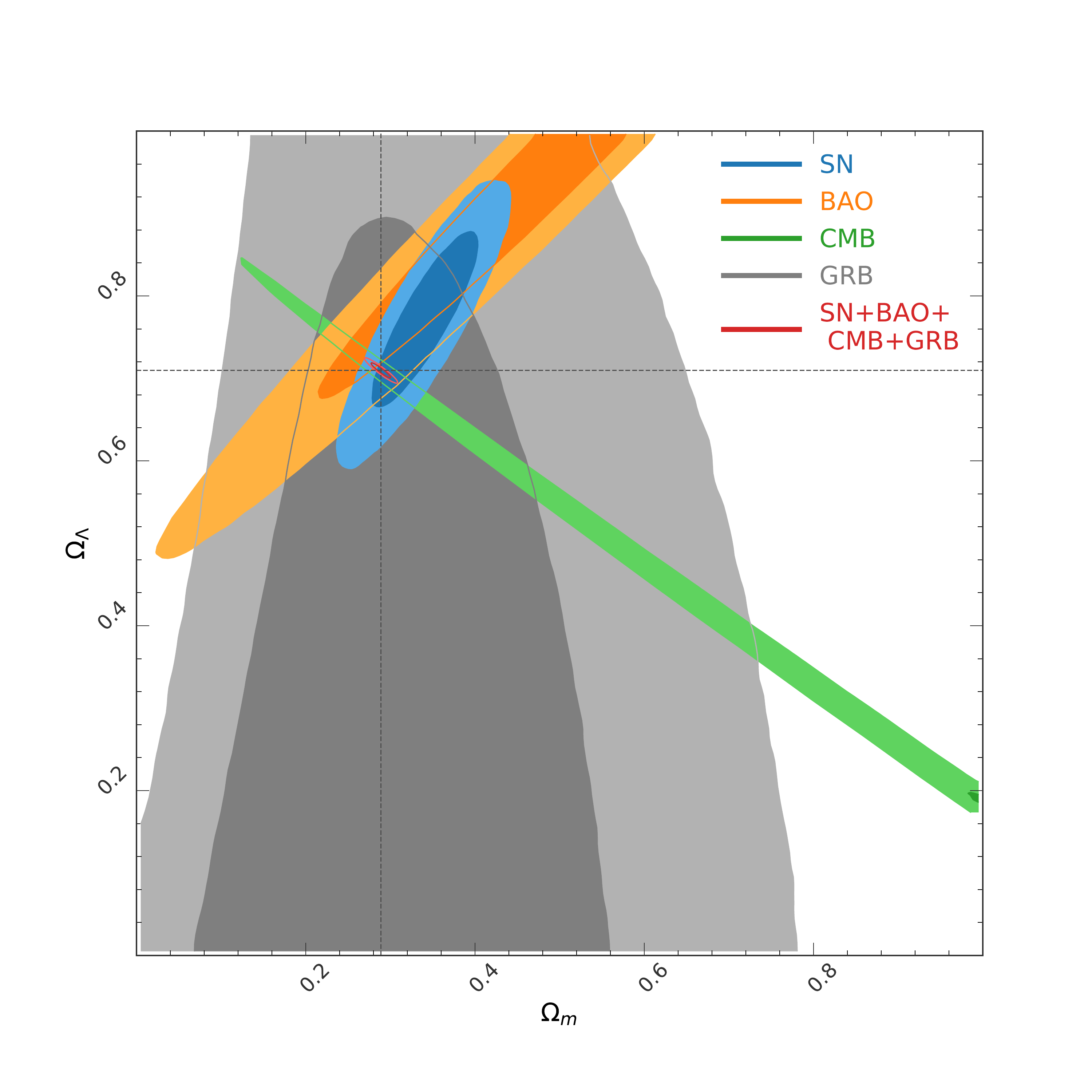}
	\caption{Constraints on $\om-\oL$ for the non-flat $\Lambda$CDM model
    (68\% and 95\% confidence levels) with observational data of SNe Ia (blue),
    BAO (orange), CMB (green) and the de-evolved \LTEp correlation of GRBs (grey).
    The joint constraint from all the four probes is shown as the
    inner red contours. The dashed lines show the most probable value of
    $\om$ and $\oL$ derived by using the four probes, i.e., $\om=0.289\pm0.008$,
    $\oL=0.710\pm0.006$}
	\label{fig:10}
\end{figure}

\begin{figure}[htbp]
	\centering
	\includegraphics[width=0.8\linewidth]{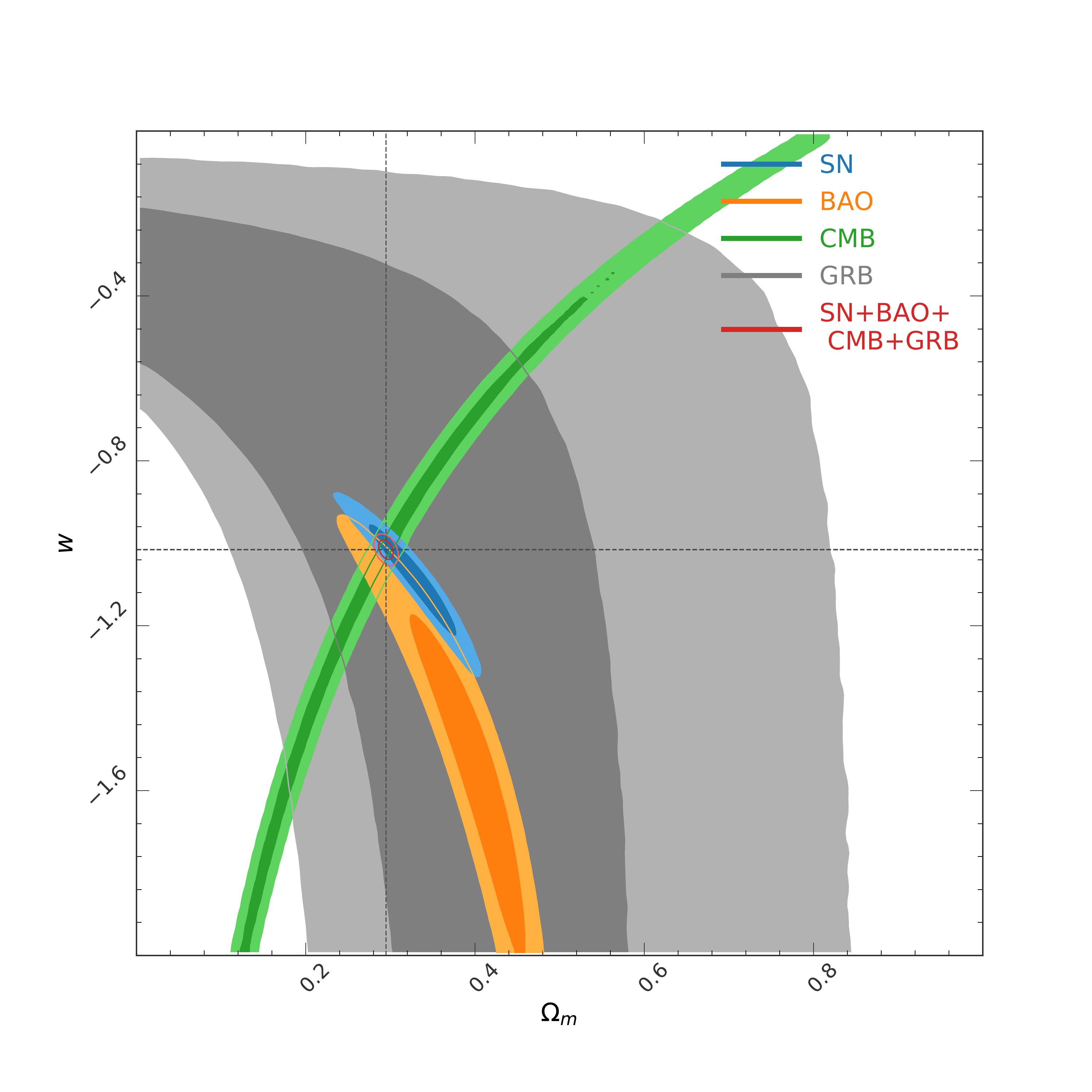}
	\caption{Constraints on $\om-w$ for the flat $w$CDM model
    (68\% and 95\% confidence levels) with observational data of SNe Ia (blue),
    BAO (orange), CMB (green) and the de-evolved \LTEp correlation of GRBs (grey).
    The joint constraint from the four cosmological probes is shown as the inner
    red contours. The dashed lines show the most probable value of $\om$ and $w$
    derived by using the four probes, i.e., $\om=0.295\pm0.006$ and $w=-1.015\pm0.015$. }
	\label{fig:11}
\end{figure}

We have used the de-evolved \LTEp correlation to constrain the cosmological
parameters. When only this probe is engaged, we obtain a result of
$\om=0.389^{+0.202}_{-0.141} (1\sigma)$ for the flat $\Lambda$CDM model.
We see that the uncertainty is
quite large, due to the fact that the correlation is still a bit too
dispersive (see Figure~\ref{fig:6}). However, this result does not conflict
with our current understanding of the Universe.
It is also consistent with the cosmological results
derived from the Combo-relation \citep{Izzo..2015, Muccino..2021}.
Combined with other probes
such as the SNe Ia \citep{Betoule..2014,Pantheon..2018,DES-SN..2019},
CMB \citep{Planck..2016,Planck..2018}, and BAO \citep{SDSS-III..2017},
we obtain  $\om=0.289\pm0.008$ and $\oL=0.710\pm0.006$ for the non-flat
$\Lambda$CDM model. The result is plotted in Figure~\ref{fig:10}. Although
the \LTEp correlation can only play a very limited role in the practice
currently, it may provide useful help when the sample size increases
significantly and when the correlation is more accurately calibrated in the
future.

For the flat $w$CDM model, we have also applied the above method to derive the
cosmological parameters. When only the de-evolved \LTEp correlation is used,
we get $\om=0.369^{+0.217}_{-0.191}$ and $w=-0.966^{+0.513}_{-0.678} (1\sigma)$,
where the uncertainties are quite large. By combining all the four probes,
the best-fit result is $\om=0.295\pm0.006$ and $w=-1.015\pm0.015$.
This result is consistent with the cosmological-constant model for dark energy.
The fitting results are shown in Figure \ref{fig:11}, and a summary of all
our cosmology results are listed in Table \ref{tab:4}.

\section{Discussion and conclusions} \label{sec:concl}

According to the \LTE correlation \citep{Xu..2012, Tang..2019, Zhao..2019},
the end time of the plateau phase ($T_a$) and the corresponding X-ray luminosity
at that moment ($L_X$) are closely related with the isotropic $\gamma$-ray
energy release during the prompt burst phase ($\Eiso$). It has been expected
that this relation may act as a standard candle to provide useful constraints on
cosmological parameters. However, in this study, it is shown that there is a
clear redshift evolution in the \LTE correlation, which makes it not a
satisfactory standard candle. Additionally, we notice that while both $L_X$
and $\Eiso$ are proportional to the luminosity distance, their power-law indices
in the correlation function are almost identical. It further means that the
\LTE correlation is very insensitive to cosmological parameters.
Considering that $\Eiso$ is closely related to the peak spectral
energy ($E_{\rm p}$), we went further to build a new three parameter correlation,
namely the \LTEp correlation, which connects the three parameters of $L_X$, $T_a$,
and $E_{\rm p}$ as $L_{X} \propto T_a^{-1.08 \pm 0.08}\Epi^{0.76 \pm 0.14}$.
The new \LTEp correlation is derived from a sample of
121 long GRBs with a plateau phase in the X-ray afterglow lightcurve.
It is largely consistent with the four-parameter
Combo-relation of $L_X \propto (T_a/|1+\alpha|)^{-1} E_p^{0.84 \pm 0.08}$
\citep{Izzo..2015, Muccino..2021}.
Our \LTEp correlation also suffers from the redshift evolution, but it can be
easily corrected for by dividing $L_X$ with $(1+z)^{1.78}$. It is found that
the best-fit de-evolved \LTEp correlation reads $L_{X,\rm{cal}}
\propto T_a^{-0.89}\Epi^{0.50}$. As a test, we have tried to combine this correlation
with other probes to constrain the cosmological parameters. Although
the \LTEp correlation can only improve the parameters marginally at
current stage, it is expected that encouraging results may be available
from this new probe when the sample size increases significantly in the
future.

Our results are consistent with the normal vision that both high-redshift
Universe and low-redshift Universe follow the same pattern.
However, some studies have shown a deviation from the $\Lambda$CDM model
when using high-redshift cosmological probes like quasars and GRBs
\citep{Lusso..2019,Amati..2019,Risaliti..2019,Demianski..2019,Luongo..2020a, Luongo..2020b}.
We want to point out that this tension may not necessarily lead to new
fundamental physics beyond the standard model, but may be related with
the redshift evolution in the empirical luminosity relations of GRBs
and quasars that are used as the candles, and with the calibration method
using SNe Ia.  The cosmological results at high redshift will be distorted
when a kind of ``standard candles'' are calibrated with low-redshift sample and then
directly extended to high redshift if there exists a redshift evolution in the
candles themselves. In our studies, we have tried to compensate the redshift
evolution in the \LTEp correlation as far as possible.

It should also be mentioned that the de-evolved \LTEp correlation derived here
is still not a very compact relation. It has a relatively large dispersion
of $\ext\sim0.5$. The reason may be due to the fact that the three parameters
involved are not measured accurately enough. First, the $\Epi$ of different
GRBs are extrapolated from the observational data of different detectors. There may
be systematic bias in the parameter. Second, for a slowly broken X-ray lightcurve, the
determination of the breaking point, i.e. $T_{a}$, is difficult and may subject to
large uncertainties. Thirdly, the X-ray luminosities ($L_{X}$) in our sample have been
derived from the XRT/\emph{Swift} observations. However, we know that XRT has a
relatively narrow passband, it may bring in a relatively large uncertainty in $L_{X}$.
In the future, we hope the three parameters of $L_{X}$, $T_{a}$, and $\Epi$ can
be accurately determined for an even larger sample. Then even the \LTEp correlation
itself may be able to give an accurate constraint on the cosmology model.

\acknowledgments
We would like to thank the anonymous referee for helpful suggestions that lead to an overall
improvement of this study. We also thank Zuo-Lin Tu, Guo-Qiang Zhang, Jian-Ping Hu, Hai Yu, and Long Li for stimulating discussion.
This work was supported by National SKA Program of China No. 2020SKA0120300,
by the National Natural Science Foundation of China (Grant Nos. 11873030, 12041306, U1938201, 11903019, and 11833003),
and by the Strategic Priority Research Program of the Chinese
Academy of Sciences (``multi-waveband Gravitational-Wave Universe'', Grant No. XDB23040000).

\nocite{*}
\bibliographystyle{aasjournal}
\bibliography{bibtex}

\begin{startlongtable}
	\begin{deluxetable}{lccccccc}
		\tablecaption{The sample of 121 GRBs used for the \LTEp study.  \label{tab:2}}
		\tablehead{
			\colhead{GRB Name} & \colhead{$z^{a}$} & \colhead{$\log(L_{X})^{a}$} & \colhead{$\log(L_{X,\rm{cal}})^{b}$} & \colhead{$\log(T_{a})^{a}$} & \colhead{$\log(\Epi)^{c}$} & \colhead{Detector} & \colhead{Refs$^{c}$}\\
			\colhead{} & \colhead{} & \colhead{($10^{47}$erg/s)} & \colhead{($10^{47}$erg/s)} & \colhead{($10^{3}$s)} & \colhead{(keV)} & \colhead{} & \colhead{}
		}
		\startdata
		GRB 050315 & 1.949 & -0.3 $\pm$ 0.24 & -1.13 $\pm$ 0.27 & 1.53 $\pm$ 0.16 & 2.07 $\pm$ 0.11 & Swift & 6 \\
		GRB 050319 & 3.24 & 0.86 $\pm$ 0.22 & -0.26 $\pm$ 0.28 & 0.53 $\pm$ 0.25 & 2.47 $\pm$ 0.34 & Swift & 6 \\
		GRB 050401 & 2.9 & 1.38 $\pm$ 0.18 & 0.33 $\pm$ 0.25 & 0.21 $\pm$ 0.14 & 2.61 $\pm$ 0.05 & Konus-Wind & 1 \\
		GRB 050416A & 0.6535 & -0.81 $\pm$ 0.17 & -1.2 $\pm$ 0.18 & 0.2 $\pm$ 0.2 & 1.4 $\pm$ 0.07 & Swift & 1 \\
		GRB 050505 & 4.27 & 1.29 $\pm$ 0.15 & 0.01 $\pm$ 0.26 & 0.37 $\pm$ 0.08 & 2.82 $\pm$ 0.16 & Swift & 1 \\
		GRB 050814 & 5.3 & 0.52 $\pm$ 0.35 & -0.9 $\pm$ 0.42 & 0.66 $\pm$ 0.3 & 2.53 $\pm$ 0.06 & Swift & 1 \\
		GRB 050922C & 2.199 & 1.96 $\pm$ 0.19 & 1.06 $\pm$ 0.24 & -0.92 $\pm$ 0.14 & 2.62 $\pm$ 0.12 & HETE-2 & 1 \\
		GRB 051016B & 0.9364 & -1.13 $\pm$ 0.29 & -1.64 $\pm$ 0.3 & 0.95 $\pm$ 0.19 & 1.73 $\pm$ 0.2 & Swift & 6 \\
		GRB 051109A & 2.346 & 1.49 $\pm$ 0.15 & 0.56 $\pm$ 0.21 & -0.19 $\pm$ 0.11 & 2.76 $\pm$ 0.16 & Konus-Wind & 1 \\
		GRB 060115 & 3.53 & 0.33 $\pm$ 0.22 & -0.83 $\pm$ 0.29 & 0.38 $\pm$ 0.19 & 2.45 $\pm$ 0.05 & Swift & 1 \\
		GRB 060116 & 4 & 1.2 $\pm$ 0.22 & -0.04 $\pm$ 0.3 & -0.66 $\pm$ 0.27 & 2.81 $\pm$ 0.18 & Swift & 2 \\
		GRB 060206 & 4.05 & 1.61 $\pm$ 0.22 & 0.36 $\pm$ 0.3 & -0.49 $\pm$ 0.2 & 2.6 $\pm$ 0.05 & Swift & 1 \\
		GRB 060210 & 3.91 & 1.58 $\pm$ 0.17 & 0.35 $\pm$ 0.26 & 0.24 $\pm$ 0.09 & 2.76 $\pm$ 0.14 & Swift & 1 \\
		GRB 060502A & 1.51 & 0.03 $\pm$ 0.2 & -0.69 $\pm$ 0.23 & 0.49 $\pm$ 0.17 & 2.51 $\pm$ 0.13 & Konus-Wind & 1 \\
		GRB 060522 & 5.11 & 1.75 $\pm$ 0.25 & 0.35 $\pm$ 0.34 & -0.89 $\pm$ 0.26 & 2.63 $\pm$ 0.08 & Swift & 1 \\
		GRB 060526 & 3.21 & 0.27 $\pm$ 0.29 & -0.84 $\pm$ 0.35 & 0.63 $\pm$ 0.21 & 2.02 $\pm$ 0.09 & Swift & 1 \\
		GRB 060605 & 3.8 & 0.9 $\pm$ 0.21 & -0.31 $\pm$ 0.29 & 0.3 $\pm$ 0.12 & 2.69 $\pm$ 0.22 & Swift & 1 \\
		GRB 060607A & 3.082 & 1.41 $\pm$ 0.02 & 0.33 $\pm$ 0.18 & 0.5 $\pm$ 0.01 & 2.76 $\pm$ 0.15 & Swift & 1 \\
		GRB 060614 & 0.13 & -2.79 $\pm$ 0.11 & -2.89 $\pm$ 0.11 & 1.69 $\pm$ 0.04 & 2.53 $\pm$ 0.23 & Swift & 1 \\
		GRB 060707 & 3.43 & 0.59 $\pm$ 0.39 & -0.56 $\pm$ 0.43 & 0.47 $\pm$ 0.33 & 2.45 $\pm$ 0.04 & Swift & 1 \\
		GRB 060708 & 2.3 & 0.6 $\pm$ 0.19 & -0.32 $\pm$ 0.24 & 0.07 $\pm$ 0.14 & 2.47 $\pm$ 0.22 & Swift & 2 \\
		GRB 060714 & 2.71 & 0.97 $\pm$ 0.13 & -0.05 $\pm$ 0.21 & -0.01 $\pm$ 0.1 & 2.37 $\pm$ 0.2 & Swift & 1 \\
		GRB 060729 & 0.54 & -0.9 $\pm$ 0.04 & -1.23 $\pm$ 0.07 & 1.62 $\pm$ 0.02 & 2.49 $\pm$ 0.1 & Swift & 2 \\
		GRB 060814 & 0.84 & -0.38 $\pm$ 0.17 & -0.86 $\pm$ 0.19 & 0.74 $\pm$ 0.09 & 2.76 $\pm$ 0.22 & Konus-Wind & 5 \\
		GRB 060906 & 3.685 & 0.46 $\pm$ 0.16 & -0.73 $\pm$ 0.25 & 0.43 $\pm$ 0.1 & 2.32 $\pm$ 0.09 & Swift & 1 \\
		GRB 060908 & 1.8836 & 1.37 $\pm$ 0.2 & 0.55 $\pm$ 0.24 & -0.7 $\pm$ 0.21 & 2.71 $\pm$ 0.09 & Swift & 1 \\
		GRB 061021 & 0.3463 & -1.11 $\pm$ 0.13 & -1.34 $\pm$ 0.14 & 0.61 $\pm$ 0.08 & 2.85 $\pm$ 0.18 & Konus-Wind & 1 \\
		GRB 061121 & 1.314 & 0.9 $\pm$ 0.14 & 0.26 $\pm$ 0.17 & 0.22 $\pm$ 0.07 & 3.15 $\pm$ 0.03 & Konus-Wind & 1 \\
		GRB 061222A & 2.088 & 1.09 $\pm$ 0.33 & 0.22 $\pm$ 0.36 & 0.42 $\pm$ 0.19 & 2.96 $\pm$ 0.04 & Konus-Wind & 1 \\
		GRB 070129 & 2.3384 & 0.07 $\pm$ 0.15 & -0.86 $\pm$ 0.21 & 0.84 $\pm$ 0.13 & 2.16 $\pm$ 0.12 & Swift & 2 \\
		GRB 070208 & 1.165 & -0.59 $\pm$ 0.22 & -1.19 $\pm$ 0.24 & 0.59 $\pm$ 0.23 & 2.05 $\pm$ 0.01 & Swift & 2 \\
		GRB 070306 & 1.497 & 0.19 $\pm$ 0.07 & -0.52 $\pm$ 0.14 & 1.09 $\pm$ 0.04 & 2.36 $\pm$ 0.08 & Swift & 6 \\
		GRB 070508 & 0.82 & 0.92 $\pm$ 0.16 & 0.46 $\pm$ 0.18 & -0.06 $\pm$ 0.24 & 2.53 $\pm$ 0.01 & Swift & 1 \\
		GRB 070714B & 0.92 & -0.13 $\pm$ 0.26 & -0.64 $\pm$ 0.28 & -0.13 $\pm$ 0.22 & 3.33 $\pm$ 0.24 & Swift & 5 \\
		GRB 071020 & 2.145 & 0.58 $\pm$ 0.29 & -0.3 $\pm$ 0.33 & 0.5 $\pm$ 0.2 & 3.01 $\pm$ 0.05 & Konus-Wind & 1 \\
		GRB 080516 & 3.2 & 0.79 $\pm$ 0.22 & -0.32 $\pm$ 0.28 & -0.02 $\pm$ 0.22 & 2.51 $\pm$ 0 & Swift & 2 \\
		GRB 080603B & 2.69 & 1.8 $\pm$ 0.19 & 0.79 $\pm$ 0.25 & -0.64 $\pm$ 0.25 & 2.57 $\pm$ 0.13 & Konus-Wind & 1 \\
		GRB 080605 & 1.6398 & 1.51 $\pm$ 0.21 & 0.76 $\pm$ 0.24 & -0.22 $\pm$ 0.13 & 2.84 $\pm$ 0.02 & Konus-Wind & 1 \\
		GRB 080721 & 2.602 & 2.74 $\pm$ 0.2 & 1.75 $\pm$ 0.26 & -0.51 $\pm$ 0.1 & 3.25 $\pm$ 0.04 & Konus-Wind & 1 \\
		GRB 080810 & 3.35 & 2 $\pm$ 0.18 & 0.86 $\pm$ 0.26 & -0.43 $\pm$ 0.12 & 3.17 $\pm$ 0.05 & Swift & 1 \\
		GRB 080905B & 2.374 & 1.42 $\pm$ 0.15 & 0.48 $\pm$ 0.22 & 0 $\pm$ 0.11 & 2.79 $\pm$ 0.12 & Fermi & 3 \\
		GRB 081007 & 0.5295 & -0.93 $\pm$ 0.21 & -1.26 $\pm$ 0.22 & 0.41 $\pm$ 0.18 & 1.79 $\pm$ 0.11 & Swift & 1 \\
		GRB 081008 & 1.9685 & 0.28 $\pm$ 0.26 & -0.57 $\pm$ 0.29 & 0.48 $\pm$ 0.15 & 2.42 $\pm$ 0.09 & Fermi & 4 \\
		GRB 081221 & 2.26 & 2.38 $\pm$ 0.1 & 1.47 $\pm$ 0.18 & -0.72 $\pm$ 0.07 & 2.42 $\pm$ 0.01 & Konus-Wind & 1 \\
		GRB 090113 & 1.7493 & 1.3 $\pm$ 0.11 & 0.51 $\pm$ 0.17 & -0.71 $\pm$ 0.11 & 2.59 $\pm$ 0.09 & Fermi & 3 \\
		GRB 090205 & 4.7 & 0.87 $\pm$ 0.3 & -0.47 $\pm$ 0.37 & -0.03 $\pm$ 0.22 & 2.33 $\pm$ 0.15 & Swift & 4 \\
		GRB 090407 & 1.4485 & -0.84 $\pm$ 0.22 & -1.53 $\pm$ 0.25 & 1.56 $\pm$ 0.15 & 2.88 $\pm$ 0 & Swift & 2 \\
		GRB 090418A & 1.608 & 1.09 $\pm$ 0.18 & 0.35 $\pm$ 0.22 & -0.07 $\pm$ 0.09 & 3.2 $\pm$ 0.11 & Swift & 1 \\
		GRB 090423 & 8 & 1.63 $\pm$ 0.12 & -0.07 $\pm$ 0.3 & -0.3 $\pm$ 0.09 & 2.61 $\pm$ 0.11 & Fermi & 4 \\
		GRB 090516 & 4.109 & 1.25 $\pm$ 0.14 & -0.01 $\pm$ 0.25 & 0.32 $\pm$ 0.07 & 2.98 $\pm$ 0.18 & Swift & 1 \\
		GRB 090519 & 3.9 & 0.12 $\pm$ 0.28 & -1.11 $\pm$ 0.34 & -0.24 $\pm$ 0.38 & 3.9 $\pm$ 0.15 & Fermi & 3 \\
		GRB 090530 & 1.266 & -0.29 $\pm$ 0.2 & -0.93 $\pm$ 0.23 & 0.23 $\pm$ 0.21 & 2.32 $\pm$ 0.16 & Swift & 2 \\
		GRB 090618 & 0.54 & 0.2 $\pm$ 0.24 & -0.14 $\pm$ 0.24 & 0.5 $\pm$ 0.19 & 2.46 $\pm$ 0.01 & Konus-Wind & 1 \\
		GRB 090927 & 1.37 & -0.5 $\pm$ 0.16 & -1.17 $\pm$ 0.19 & 0.6 $\pm$ 0.13 & 2.67 $\pm$ 0.15 & Fermi & 3 \\
		GRB 091018 & 0.971 & 0.84 $\pm$ 0.16 & 0.31 $\pm$ 0.18 & -0.53 $\pm$ 0.15 & 1.74 $\pm$ 0.21 & Konus-Wind & 1 \\
		GRB 091029 & 2.752 & 0.66 $\pm$ 0.13 & -0.37 $\pm$ 0.21 & 0.38 $\pm$ 0.09 & 2.36 $\pm$ 0.12 & Swift & 1 \\
		GRB 091127 & 0.49 & -0.23 $\pm$ 0.22 & -0.53 $\pm$ 0.23 & 1.03 $\pm$ 0.51 & 1.73 $\pm$ 0.02 & Swift & 1 \\
		GRB 091208B & 1.0633 & 0.58 $\pm$ 0.23 & 0.02 $\pm$ 0.25 & -0.4 $\pm$ 0.13 & 2.39 $\pm$ 0.04 & Fermi & 4 \\
		GRB 100302A & 4.813 & 0.31 $\pm$ 0.19 & -1.05 $\pm$ 0.29 & 0.22 $\pm$ 0.21 & 2.72 $\pm$ 0.17 & Swift & 2 \\
		GRB 100418A & 0.6235 & -1.85 $\pm$ 0.1 & -2.22 $\pm$ 0.12 & 1.69 $\pm$ 0.11 & 1.67 $\pm$ 0.03 & Swift & 1 \\
		GRB 100424A & 2.465 & 2.51 $\pm$ 0.15 & 1.55 $\pm$ 0.22 & -1.08 $\pm$ 0.05 & 2.52 $\pm$ 0.07 & Swift & 2 \\
		GRB 100615A & 1.398 & 0.38 $\pm$ 0.17 & -0.3 $\pm$ 0.21 & 0.86 $\pm$ 0.18 & 2.11 $\pm$ 0.06 & Fermi & 3 \\
		GRB 100621A & 0.542 & -0.86 $\pm$ 0.23 & -1.2 $\pm$ 0.24 & 1.2 $\pm$ 0.18 & 2.21 $\pm$ 0.03 & Konus-Wind & 1 \\
		GRB 100704A & 3.6 & 1.13 $\pm$ 0.19 & -0.05 $\pm$ 0.27 & 0.36 $\pm$ 0.14 & 2.91 $\pm$ 0.08 & Fermi & 3 \\
		GRB 100814A & 1.44 & -0.47 $\pm$ 0.27 & -1.16 $\pm$ 0.29 & 1.84 $\pm$ 0.07 & 2.49 $\pm$ 0.04 & Konus-Wind & 1 \\
		GRB 100902A & 4.5 & -0.33 $\pm$ 0.13 & -1.65 $\pm$ 0.25 & 2.18 $\pm$ 0.07 & 2.67 $\pm$ 0 & Swift & 2 \\
		GRB 100906A & 1.727 & 0.3 $\pm$ 0.31 & -0.48 $\pm$ 0.33 & 0.7 $\pm$ 0.16 & 2.46 $\pm$ 0.08 & Swift & 1 \\
		GRB 101219B & 0.5519 & -2.08 $\pm$ 0.17 & -2.42 $\pm$ 0.18 & 1.22 $\pm$ 0.25 & 2.04 $\pm$ 0.05 & Swift & 1 \\
		GRB 110213A & 1.46 & 1.42 $\pm$ 0.23 & 0.72 $\pm$ 0.26 & 0.01 $\pm$ 0.07 & 2.38 $\pm$ 0.03 & Swift & 1 \\
		GRB 110715A & 0.82 & 1.32 $\pm$ 0.09 & 0.85 $\pm$ 0.11 & -0.84 $\pm$ 0.09 & 2.34 $\pm$ 0.03 & Konus-Wind & 1 \\
		GRB 111008A & 5 & 1.79 $\pm$ 0.13 & 0.41 $\pm$ 0.26 & -0.06 $\pm$ 0.08 & 2.8 $\pm$ 0.11 & Konus-Wind & 1 \\
		GRB 111123A & 3.1516 & 0.51 $\pm$ 0.2 & -0.59 $\pm$ 0.27 & 0.7 $\pm$ 0.15 & 3.6 $\pm$ 0.05 & Swift & 2 \\
		GRB 111209A & 0.677 & -1.57 $\pm$ 0.49 & -1.97 $\pm$ 0.5 & 1.5 $\pm$ 0.06 & 2.72 $\pm$ 0.07 & Swift & 1 \\
		GRB 111228A & 0.7163 & -0.4 $\pm$ 0.12 & -0.82 $\pm$ 0.14 & 0.77 $\pm$ 0.09 & 1.87 $\pm$ 0.26 & Konus-Wind & 1 \\
		GRB 111229A & 1.3805 & -0.15 $\pm$ 0.23 & -0.82 $\pm$ 0.25 & 0.57 $\pm$ 0.09 & 2.84 $\pm$ 0 & Swift & 2 \\
		GRB 120118B & 2.943 & 0.79 $\pm$ 0.17 & -0.27 $\pm$ 0.24 & 0.04 $\pm$ 0.22 & 2.23 $\pm$ 0.03 & Fermi & 3 \\
		GRB 120326A & 1.798 & 0.26 $\pm$ 0.05 & -0.53 $\pm$ 0.14 & 1.26 $\pm$ 0.02 & 2.06 $\pm$ 0.07 & Swift & 1 \\
		GRB 120327A & 2.813 & 1.46 $\pm$ 0.19 & 0.43 $\pm$ 0.25 & -0.4 $\pm$ 0.12 & 2.92 $\pm$ 0.17 & Swift & 2 \\
		GRB 120404A & 2.876 & 0.91 $\pm$ 0.17 & -0.14 $\pm$ 0.24 & -0.04 $\pm$ 0.11 & 2.94 $\pm$ 0 & Swift & 2 \\
		GRB 120521C & 6 & 0.31 $\pm$ 0.24 & -1.19 $\pm$ 0.35 & 0.51 $\pm$ 0.14 & 2.86 $\pm$ 0.19 & Swift & 2 \\
		GRB 120712A & 4.1745 & -0.55 $\pm$ 0.29 & -1.82 $\pm$ 0.35 & 1.65 $\pm$ 0.17 & 2.81 $\pm$ 0.09 & Fermi & 1 \\
		GRB 120802A & 3.796 & 0.72 $\pm$ 0.21 & -0.49 $\pm$ 0.29 & 0.18 $\pm$ 0.3 & 2.44 $\pm$ 0.15 & Swift & 4 \\
		GRB 120811C & 2.671 & 1.16 $\pm$ 0.16 & 0.16 $\pm$ 0.23 & -0.14 $\pm$ 0.18 & 2.3 $\pm$ 0.04 & Fermi & 4 \\
		GRB 120922A & 3.1 & 1.3 $\pm$ 0.12 & 0.21 $\pm$ 0.21 & -0.27 $\pm$ 0.14 & 2.17 $\pm$ 0.05 & Fermi & 3 \\
		GRB 121128A & 2.2 & 1.58 $\pm$ 0.2 & 0.68 $\pm$ 0.25 & -0.42 $\pm$ 0.1 & 2.39 $\pm$ 0.02 & Konus-Wind & 1 \\
		GRB 121211A & 1.023 & -0.46 $\pm$ 0.17 & -1.01 $\pm$ 0.19 & 0.64 $\pm$ 0.16 & 2.31 $\pm$ 0.06 & Fermi & 3 \\
		GRB 130408A & 3.758 & 0.6 $\pm$ 0.2 & -0.6 $\pm$ 0.28 & 0.76 $\pm$ 0.07 & 3.11 $\pm$ 0.06 & Konus-Wind & 1 \\
		GRB 130420A & 1.297 & 0.08 $\pm$ 0.17 & -0.56 $\pm$ 0.2 & 0.2 $\pm$ 0.18 & 2.12 $\pm$ 0.02 & Fermi & 1 \\
		GRB 130514A & 3.6 & 1.26 $\pm$ 0.21 & 0.08 $\pm$ 0.28 & -0.18 $\pm$ 0.2 & 2.7 $\pm$ 0.13 & Konus-Wind & 1 \\
		GRB 130606A & 5.913 & 1.3 $\pm$ 0.17 & -0.19 $\pm$ 0.3 & 0.14 $\pm$ 0.12 & 3.31 $\pm$ 0.11 & Konus-Wind & 1 \\
		GRB 130612A & 2.006 & -0.17 $\pm$ 0.21 & -1.02 $\pm$ 0.26 & 0.06 $\pm$ 0.25 & 2.04 $\pm$ 0.09 & Swift & 5 \\
		GRB 131030A & 1.293 & 1.83 $\pm$ 0.1 & 1.19 $\pm$ 0.15 & -0.75 $\pm$ 0.09 & 2.65 $\pm$ 0.01 & Konus-Wind & 1 \\
		GRB 131103A & 0.599 & -0.16 $\pm$ 0.13 & -0.53 $\pm$ 0.14 & -0.23 $\pm$ 0.13 & 2.01 $\pm$ 0.21 & Swift & 2 \\
		GRB 131105A & 1.686 & 0.64 $\pm$ 0.11 & -0.13 $\pm$ 0.17 & 0.23 $\pm$ 0.1 & 2.73 $\pm$ 0.07 & Konus-Wind & 1 \\
		GRB 140206A & 2.73 & 2.06 $\pm$ 0.13 & 1.05 $\pm$ 0.21 & -0.31 $\pm$ 0.1 & 2.58 $\pm$ 0.06 & Swift & 5 \\
		GRB 140213A & 1.2076 & -0.19 $\pm$ 0.26 & -0.8 $\pm$ 0.28 & 1.28 $\pm$ 0.18 & 2.34 $\pm$ 0.02 & Konus-Wind & 1 \\
		GRB 140304A & 5.283 & 2.85 $\pm$ 0.31 & 1.43 $\pm$ 0.38 & -0.99 $\pm$ 0.11 & 2.89 $\pm$ 0.11 & Fermi & 3 \\
		GRB 140512A & 0.725 & -0.13 $\pm$ 0.14 & -0.55 $\pm$ 0.15 & 0.99 $\pm$ 0.18 & 2.92 $\pm$ 0.09 & Konus-Wind & 1 \\
		GRB 140518A & 4.707 & 1.33 $\pm$ 0.21 & -0.02 $\pm$ 0.3 & -0.31 $\pm$ 0.12 & 2.4 $\pm$ 0.07 & Swift & 1 \\
		GRB 140629A & 2.275 & 0.95 $\pm$ 0.31 & 0.03 $\pm$ 0.35 & -0.01 $\pm$ 0.24 & 2.45 $\pm$ 0.09 & Konus-Wind & 1 \\
		GRB 140703A & 3.14 & 1.05 $\pm$ 0.27 & -0.05 $\pm$ 0.33 & 0.53 $\pm$ 0.13 & 2.96 $\pm$ 0.05 & Fermi & 1 \\
		GRB 141004A & 0.57 & -0.78 $\pm$ 0.25 & -1.13 $\pm$ 0.26 & 0.03 $\pm$ 0.18 & 1.64 $\pm$ 0.1 & Fermi & 3 \\
		GRB 141121A & 1.47 & -1.18 $\pm$ 0.21 & -1.87 $\pm$ 0.24 & 2.1 $\pm$ 0.12 & 2.29 $\pm$ 0.19 & Swift & 2 \\
		GRB 150323A & 0.593 & -1.45 $\pm$ 0.24 & -1.81 $\pm$ 0.24 & 0.84 $\pm$ 0.21 & 2.18 $\pm$ 0.04 & Konus-Wind & 1 \\
		GRB 150403A & 2.06 & 2.48 $\pm$ 0.06 & 1.61 $\pm$ 0.15 & -0.31 $\pm$ 0.03 & 3.06 $\pm$ 0.04 & Konus-Wind & 1 \\
		GRB 150424A & 3 & -0.03 $\pm$ 0.29 & -1.1 $\pm$ 0.34 & 0.91 $\pm$ 0.21 & 3.08 $\pm$ 0.02 & Konus-Wind & 1 \\
		GRB 151027A & 0.81 & 0.84 $\pm$ 0.06 & 0.38 $\pm$ 0.1 & 0.37 $\pm$ 0.03 & 2.5 $\pm$ 0.12 & Konus-Wind & 1 \\
		GRB 151027B & 4.063 & 0.7 $\pm$ 0.21 & -0.55 $\pm$ 0.29 & 0.36 $\pm$ 0.22 & 2.72 $\pm$ 0 & Swift & 2 \\
		GRB 160227A & 2.38 & 0.41 $\pm$ 0.18 & -0.53 $\pm$ 0.24 & 0.89 $\pm$ 0.14 & 2.35 $\pm$ 0.11 & Swift & 1 \\
		GRB 160804A & 0.736 & -1.08 $\pm$ 0.33 & -1.5 $\pm$ 0.33 & 0.78 $\pm$ 0.22 & 2.12 $\pm$ 0.02 & Fermi & 1 \\
		GRB 161108A & 1.159 & -1.08 $\pm$ 0.34 & -1.68 $\pm$ 0.35 & 1.29 $\pm$ 0.22 & 2.15 $\pm$ 0.22 & Swift & 2 \\
		GRB 161117A & 1.549 & 0.44 $\pm$ 0.12 & -0.29 $\pm$ 0.17 & 0.42 $\pm$ 0.11 & 2.25 $\pm$ 0.04 & Konus-Wind & 1 \\
		GRB 170113A & 1.968 & 1.08 $\pm$ 0.13 & 0.24 $\pm$ 0.19 & -0.08 $\pm$ 0.09 & 2.34 $\pm$ 0.2 & Swift & 1 \\
		GRB 170202A & 3.65 & 1.61 $\pm$ 0.14 & 0.42 $\pm$ 0.24 & -0.43 $\pm$ 0.11 & 3.06 $\pm$ 0.23 & Konus-Wind & 1 \\
		GRB 170607A & 0.557 & -0.81 $\pm$ 0.12 & -1.15 $\pm$ 0.13 & 0.89 $\pm$ 0.1 & 2.35 $\pm$ 0.03 & Fermi & 1 \\
		GRB 170705A & 2.01 & 0.68 $\pm$ 0.15 & -0.17 $\pm$ 0.21 & 0.78 $\pm$ 0.13 & 2.67 $\pm$ 0.07 & Fermi & 1 \\
		GRB 170714A & 0.793 & 0.96 $\pm$ 0.02 & 0.51 $\pm$ 0.08 & 0.89 $\pm$ 0.01 & 2.37 $\pm$ 0.06 & Swift & 2 \\
		GRB 171222A & 2.409 & -0.72 $\pm$ 0.28 & -1.67 $\pm$ 0.32 & 1.47 $\pm$ 0.36 & 2 $\pm$ 0.04 & Fermi & 5 \\
		GRB 180325A & 2.25 & 1.45 $\pm$ 0.28 & 0.54 $\pm$ 0.32 & 0.16 $\pm$ 0.26 & 3 $\pm$ 0.06 & Konus-Wind & 1 \\
		GRB 180329B & 1.998 & 0.42 $\pm$ 0.14 & -0.43 $\pm$ 0.19 & 0.26 $\pm$ 0.12 & 2.16 $\pm$ 0.08 & Swift & 1 \\
		GRB 180720B & 0.654 & 0.32 $\pm$ 0.35 & -0.07 $\pm$ 0.36 & 1.33 $\pm$ 0.18 & 3.02 $\pm$ 0.01 & Fermi & 3 \\
		\enddata
		\tablenotetext{a}{Taken from \cite{Tang..2019}}
		\tablenotetext{b}{Calculated from Equation (\ref{eq:19}).}
		\tablenotetext{c}{Values of $\log(\Epi)$ and references:
				1 - \cite{Minaev..2020};
				2 - \emph{Swift} catalog \citep{Lien..2016};
				3 - GBM/\emph{Fermi} catalog \citep{Gruber..2014,Kienlin..2014,Bhat..2016,Kienlin..2020};
				4 - \cite{Demianski..2017};
				5 - GCN circulars archive;
				6 - \cite{Wang..2020}.
			}
	\end{deluxetable}
\end{startlongtable}

\begin{table}[h!]
	\renewcommand{\thetable}{\arabic{table}}
	\centering
	\caption{Redshift evolution test of the $L_{x}-T_{a}-\Epi$ correlation.}
	\label{tab:3}
	\begin{tabular}{cccccc}
		\hline
		\hline
		\LTEp correlation   &  $ a' $     & $ b' $     & $ c' $ & $\ext$      \\
		\hline
		Full data & -1.03 $\pm$ 0.37 & -1.08 $\pm$ 0.08 & 0.76 $\pm$ 0.14 & 0.54 $\pm$ 0.04 \\
		Low-redshift sub-sample & -1.01 $\pm$ 0.57 & -0.94 $\pm$ 0.13 & 0.62 $\pm$ 0.23 & 0.62 $\pm$ 0.07 \\
		High-redshift sub-sample & 0.06 $\pm$ 0.44 & -1.01 $\pm$ 0.09 & 0.44 $\pm$ 0.16 & 0.34 $\pm$ 0.05 \\
		\\
		
		\hline
		\hline
		\LTEp correlation with a redshift evolution part  & $ a' $     & $ b' $     & $ c' $   & $ d' $  & $\ext$    \\
		\hline
		Full data & -1.06 $\pm$ 0.32 & -0.96 $\pm$ 0.07 & 0.42 $\pm$ 0.14 & 1.78 $\pm$ 0.29 & 0.47 $\pm$ 0.04 \\
		\\
		
		\hline
		\hline
		de-evolved \LTEp correlation & $ a'' $     & $ b'' $     & $ c'' $   & $\ext$     \\
		\hline
		Full data & -1.10 $\pm$ 0.32 & -0.96 $\pm$ 0.07 & 0.44 $\pm$ 0.12 & 0.46 $\pm$ 0.04 \\
		Low-redshift sub-sample & -1.36 $\pm$ 0.50 & -0.89 $\pm$ 0.11 & 0.52 $\pm$ 0.21 & 0.53 $\pm$ 0.06 \\
		High-redshift sub-sample & -1.36 $\pm$ 0.51 & -0.89 $\pm$ 0.11 & 0.53 $\pm$ 0.21 & 0.53 $\pm$ 0.06 \\
		\\
		\hline
	\end{tabular}
\end{table}		

\begin{table}[h!]
	\renewcommand{\thetable}{\arabic{table}}
	\centering
	\caption{Cosmological results with the de-evolved \LTEp correlation.}
	\label{tab:4}
	\begin{tabular}{lcc}
		\hline
		\hline
		Flat $\Lambda$CDM & $\om$   \\
		\hline
		GRB & 0.3887$^{+0.2023}_{-0.1410}$ \\
		SN+BAO+CMB & 0.2914 $\pm$ 0.0057 \\
		GRB+SN+BAO+CMB & 0.2910 $\pm$ 0.0053 \\
		\\
		
		\hline
		\hline
		Non-flat $\Lambda$CDM & $\om$ &  $\oL$   \\
		\hline
		GRB & 0.3332$^{+0.1875}_{-0.1424}$ & 0.3458$^{+0.3562}_{-0.2487}$ \\
		SN+CMB+BAO & 0.2888 $\pm$ 0.0079 & 0.7095 $\pm$ 0.0062 \\
		GRB+SN+CMB+BAO & 0.2889 $\pm$ 0.0080 & 0.7096 $\pm$ 0.0062 \\
		\\
		
		\hline
		\hline
		$w$CDM & $\om$  & $w$   \\
		\hline
		GRB & 0.3694 $^{+0.2172}_{-0.1908}$ & -0.9661 $^{+0.5132}_{-0.6778}$ \\
		SN+CMB+BAO & 0.2947 $\pm$ 0.0056 & -1.0152 $\pm$ 0.0151 \\
		GRB+SN+CMB+BAO & 0.2948 $\pm$ 0.0057 & -1.0153 $\pm$ 0.0154 \\
		\\
		\hline
	\end{tabular}
\end{table}	

\begin{deluxetable}{ccccccccc}
	\tablecaption{Covariance matrix and observed data for BAO \citep{SDSS-III..2017}.  \label{tab:5}}
	\tablehead{
		\colhead{} & \colhead{Mean} & \colhead{ $\sigma{i}$ } & \multicolumn{6}{c}{$C_{ij}^{a}$}
	}
	\startdata
	$D_{M}(0.38)$ & 1518 & 22 & 484 & 9.5304 & 295.218 & 4.6691 & 140.1664 & 2.4024 \\
	$H_{z}(0.38)$ & 81.5 & 1.9 & 9.5304 & 3.61 & 7.8797 & 1.7592 & 5.9827 & 0.9205 \\
	$D_{M}(0.51)$ & 1977 & 27 & 295.218 & 7.8797 & 729 & 11.9324 & 442.368 & 6.8664 \\
	$H_{z}(0.51)$ & 90.4 & 1.9 & 4.6691 & 1.7592 & 11.9324 & 3.61 & 9.5517 & 2.1742 \\
	$D_{M}(0.61)$ & 2283 & 32 & 140.1664 & 5.9827 & 442.368 & 9.5517 & 1024 & 16.1818 \\
	$H_{z}(0.61)$ & 97.3 & 2.1 & 2.4024 & 0.9205 & 6.8664 & 2.1742 & 16.1818 & 4.41 \\
	\enddata
	\tablenotetext{a}{Elements of the covariance matrix.}
\end{deluxetable}	

\appendix

\section{Description of the Efron \& Petrosian (1992) method}

The Efron \& Petrosian (1992) method (also known as the non-parametric $\tau$
statistical method) is often used to test the independence of variables in truncated data \citep{Yonetoku..2004,Dainotti..2013,Petrosian..2015,Yu..2015,Deng..2016}.
Here we use this method to reveal the possible redshift evolution of $L_{X}$, $T_{a}$, and $E_{\rm{p}}$ \citep{Dainotti..2013,Dainotti..2017c}.
We take the evolution form as $g(z)=(1+z)^{k_{L_{X}}}$, $f(z)=(1+z)^{k_{T_{a}}}$,
and $h(z)=(1+z)^{k_{E_{\rm{p}}}}$ for $L_{X}$, $T_{a}$, and $E_{\rm{p}}$, respectively.
Then, the redshift-independent parameters should be $L'_{X}=L_{X}/g(z)$, $T'_{a}=T_{a}/f(z)$,
and $E'_{\rm{p}}=E_{\rm{p}}/h(z)$.  Using $L_{X}$ as an example, we outline this method as follows.

First, we need to determine the threshold for truncated data.  This limit should keep a large
sample size and represent the sample itself at the same time.  We have chosen the same limit used
in \citet{Dainotti..2017c} for $L_{X}$ and $T_{a}$ ($F_{X,\rm{lim}}=1.0\times10^{-12}$ erg cm$^{2}$
and $T_{a,\rm{lim}}=309/(1+z)$ s respectively).  For $E_{\rm{p}}$, we have chosen the limit energy
as $E_{\rm{p,lim}}=1.39\times(1+z)$ keV.  Our criterion rules out 12 GRBs and remains a sample of 109 GRBs.
Figure \ref{fig:A1} below shows the distribution of the sample with black dots and the threshold
with solid curves.

\begin{figure}[htbp]
	\centering
	\subfloat{
		\includegraphics[width=0.32\linewidth]{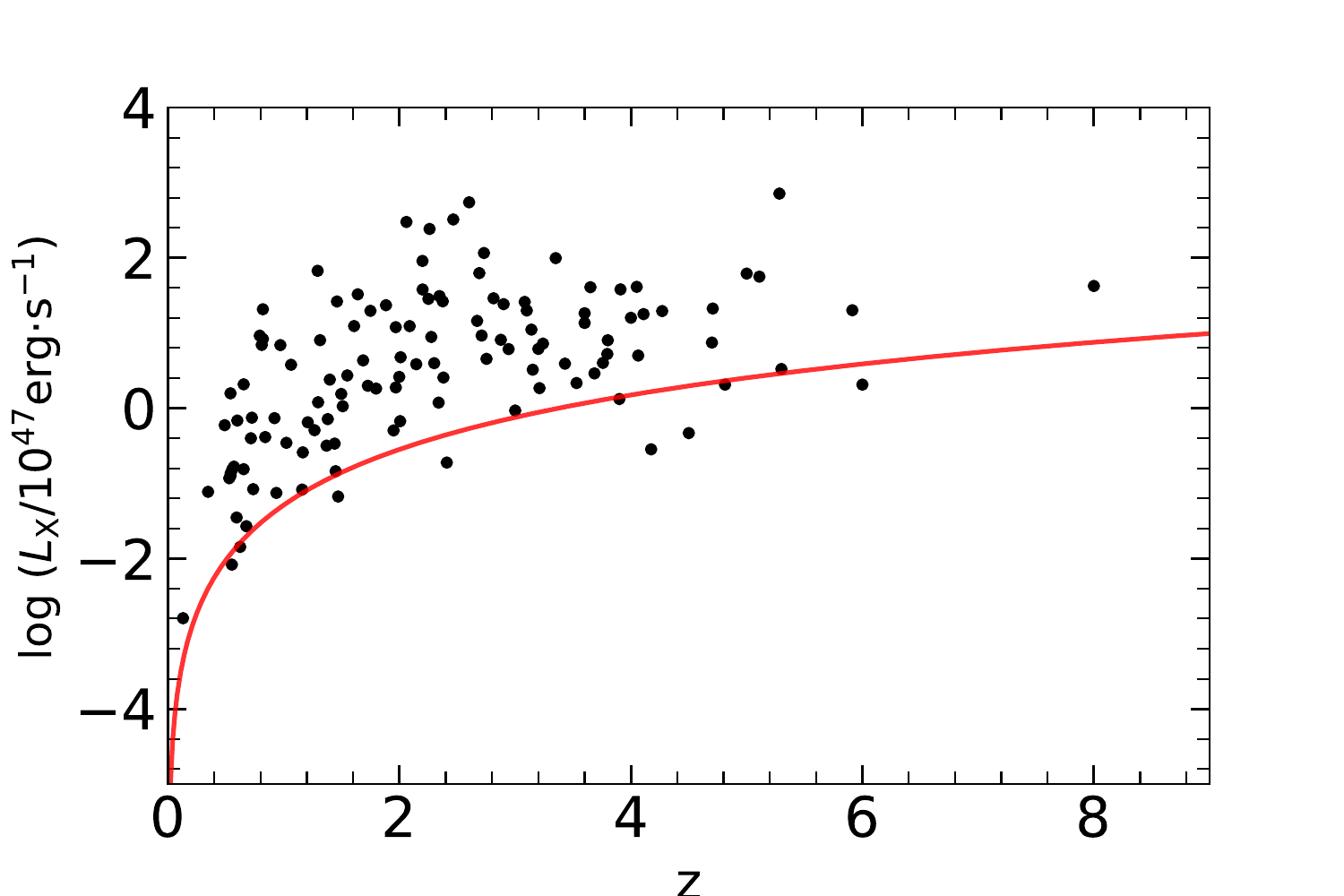}
	}
	\subfloat{
		\includegraphics[width=0.32\linewidth]{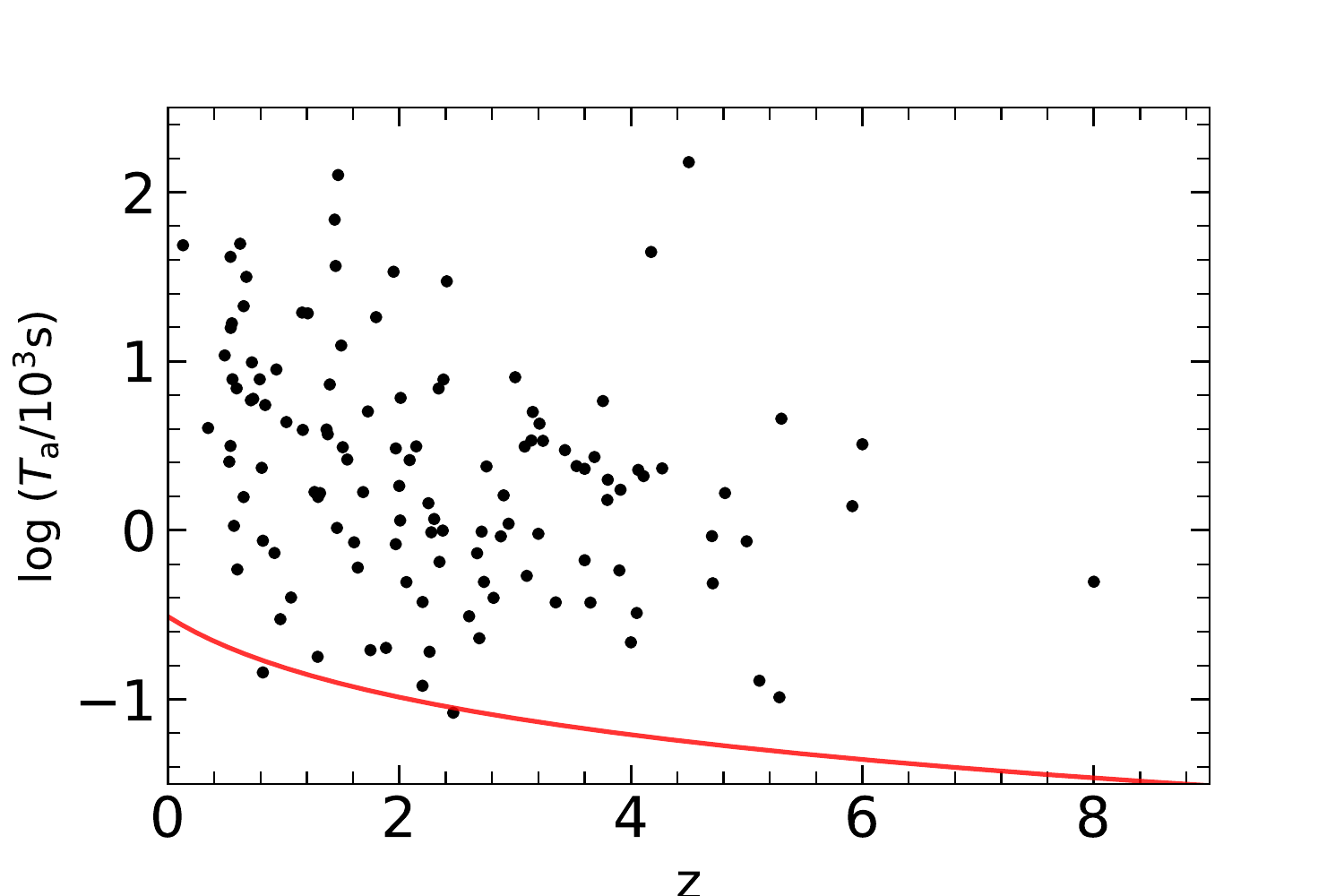}
	}
	\subfloat{
		\includegraphics[width=0.32\linewidth]{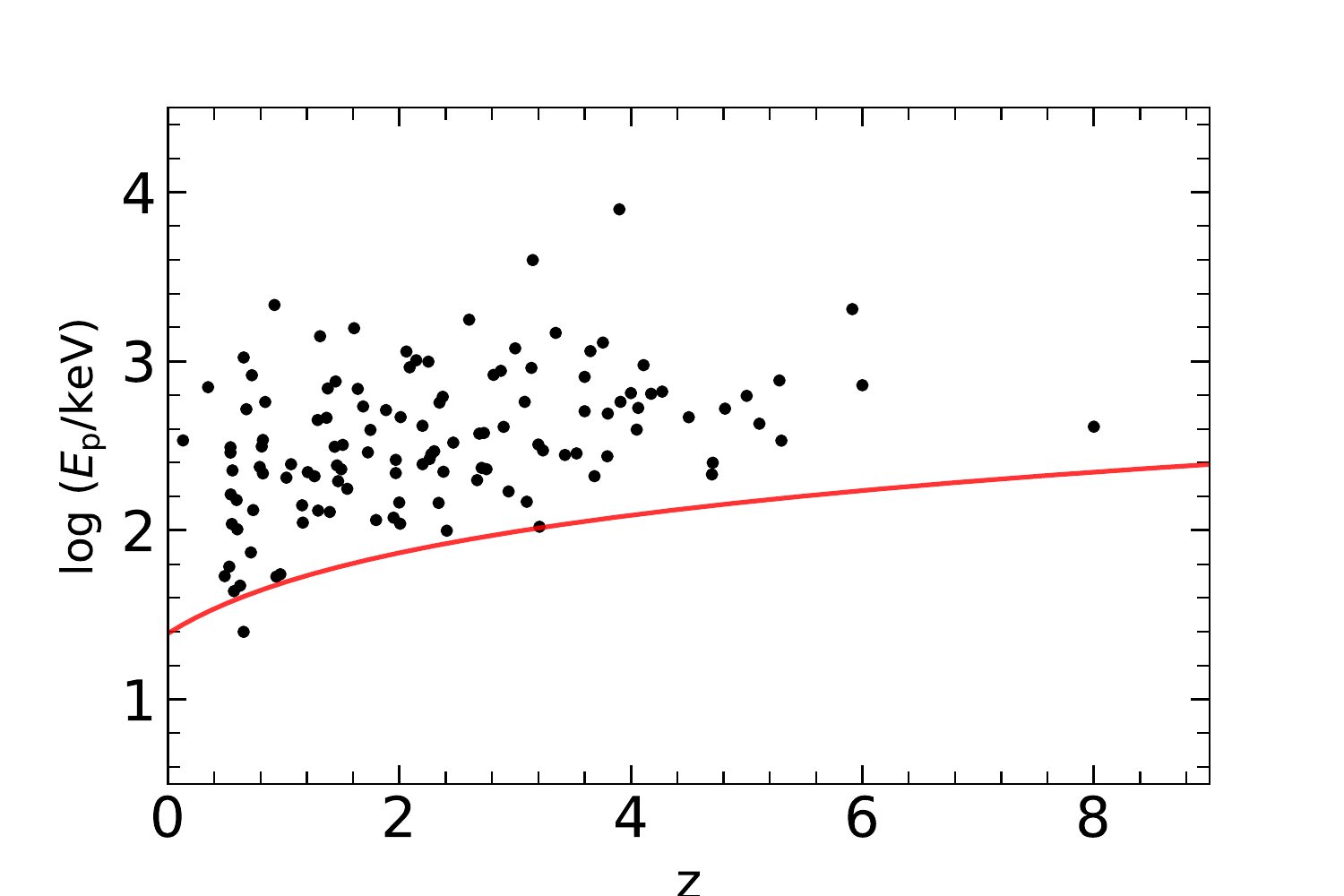}
	}
	\caption{ The distribution of the 121 GRBs on the $L_{X}-z$ plane (left panel),
the $T_{a}-z$ plane (middle panel), and the $E_{\rm{p}}-z$ plane (right panel).  The solid curves
correspond to $F_{X,\rm{lim}}=1.0\times10^{-12}$ erg cm$^{2}$, $T_{a,\rm{lim}}=309/(1+z)$ s
and $E_{\rm{p,lim}}=1.39\times(1+z)$ keV, respectively. }
	\label{fig:A1}
\end{figure}

Then, for the $i$th data $(L_{X,i},z_{i})$ in the GRB sample, we can define its associated set as

\begin{equation}
	J_{i} = \{j|L_{X,j}>L_{i}, \; \rm{and} \; z_{j}<z_{i,\rm{lim}}\},
\end{equation}
where $z_{i,\rm{lim}}$ is the redshift corresponding to the flux threshold
for a GRB of $(L_{X,i},z_{i})$.  For each data point in the sample, we can define
its associated set $J_{i}$, and the number of GRBs in the set is counted as $N_{i}$.
Also, we can get its rank $R_{i}$ in its associated set $J_{i}$ as,

\begin{equation}
	R_{i} = \rm{number}\{j \in J_{i}|z_{j} \ge z_{i}\}.
\end{equation}
The key point of this method is to determine the rank $R_{i}$ for each data point
in its associated set.  If $L_{X}$ and $z$ are independent, $R_{i}$ should be uniformly
distributed between 1 and $N_{i}$ \citep{Efron..1992}.  The test statistic $\tau$ is

\begin{equation}
	\tau = \frac{\sum_{i}(R_{i}-E_{i})}{\sqrt{\sum_{i}V_{i}}},
\end{equation}
where $E_{i}=(N_{i}+1)/2$ and $V_{i}=(N_{i}^{2}-1)/12$ are the expected mean and
the variance of $R_{i}$ respectively.  If $R_{i}$ is uniformly distributed between 1
and $N_{i}$, then the value of $\tau$ should equal to zero.  In other words, we need to
find the value of $k_{L_{X}}$ for which $\tau_{L_{X}}=0$, where $\tau_{L_{X}}$ is calculated
by using the $(L'_{X},z)$ data set.

The best $k_{L_{X}}$ we derived is $k_{L_{X}}=3.38\pm0.62$.  Using the same method,
we get $k_{T_{a}}=-1.54\pm0.30$ and $k_{E_{\rm{p}}}=0.75\pm0.25$.
The $1\sigma$ range of uncertainty is derived from $|\tau_{x}|\le1$ ($x=L_{X},T_{a},E_{\rm{p}}$).
These $k$ values represent the evolution indices for $L_{X}$, $T_{a}$, and $E_{\rm{p}}$, respectively.
In Figure \ref{fig:A2} below, for clarity, we have plotted $\tau$ versus $k$ for the three parameters.

\begin{figure}[htbp]
	\centering
	\subfloat{
		\includegraphics[width=0.32\linewidth]{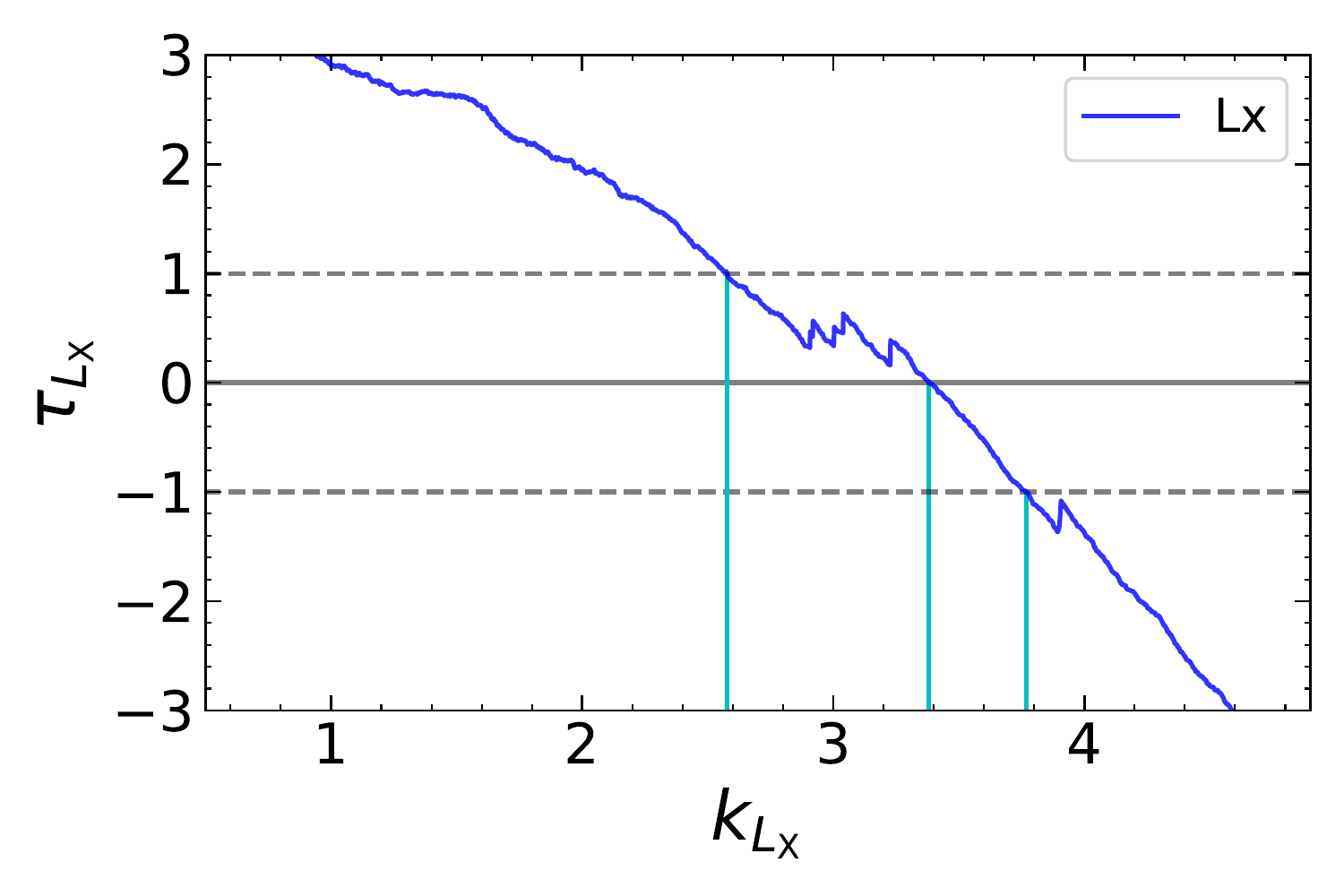}
	}
	\subfloat{
		\includegraphics[width=0.32\linewidth]{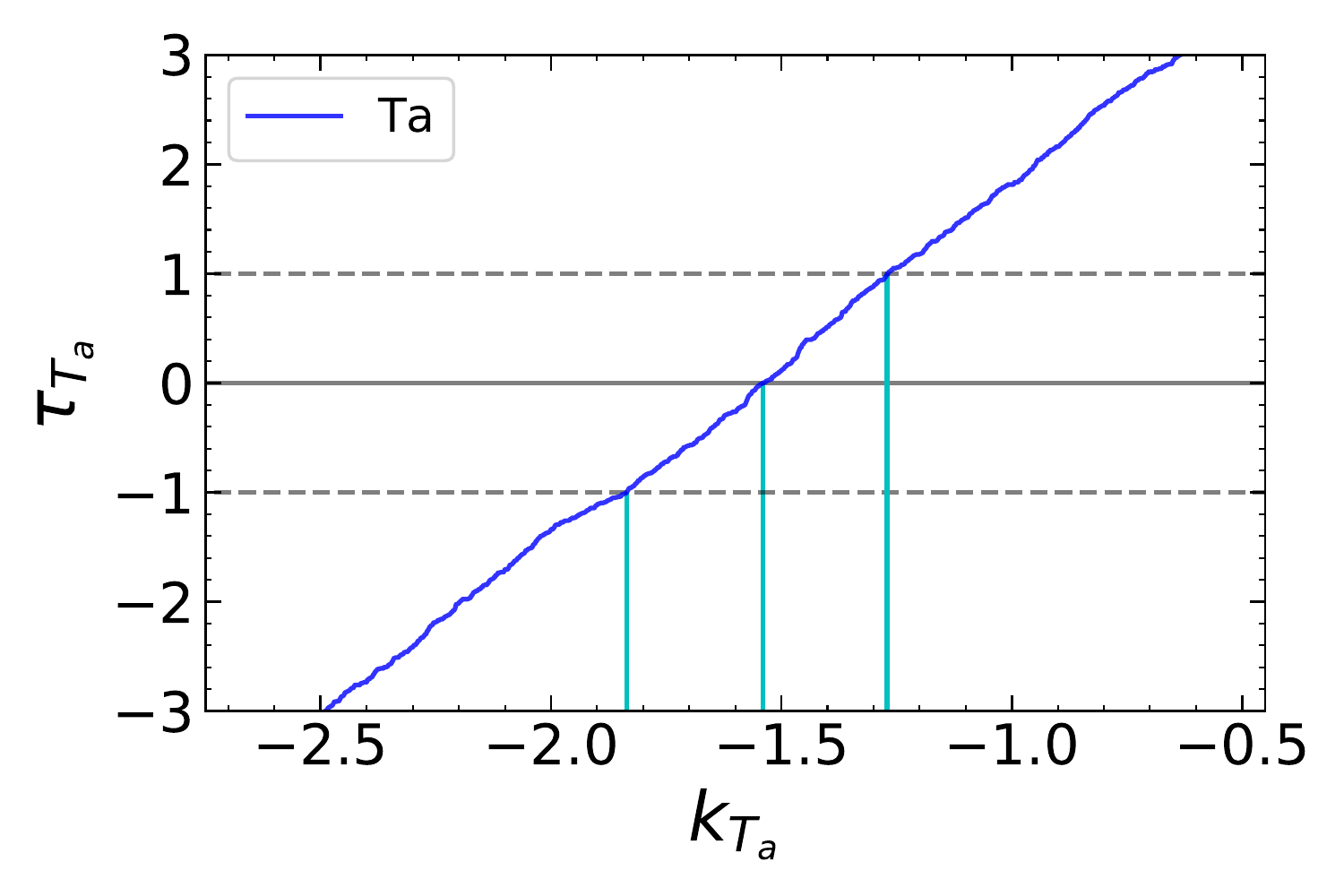}
	}
	\subfloat{
		\includegraphics[width=0.32\linewidth]{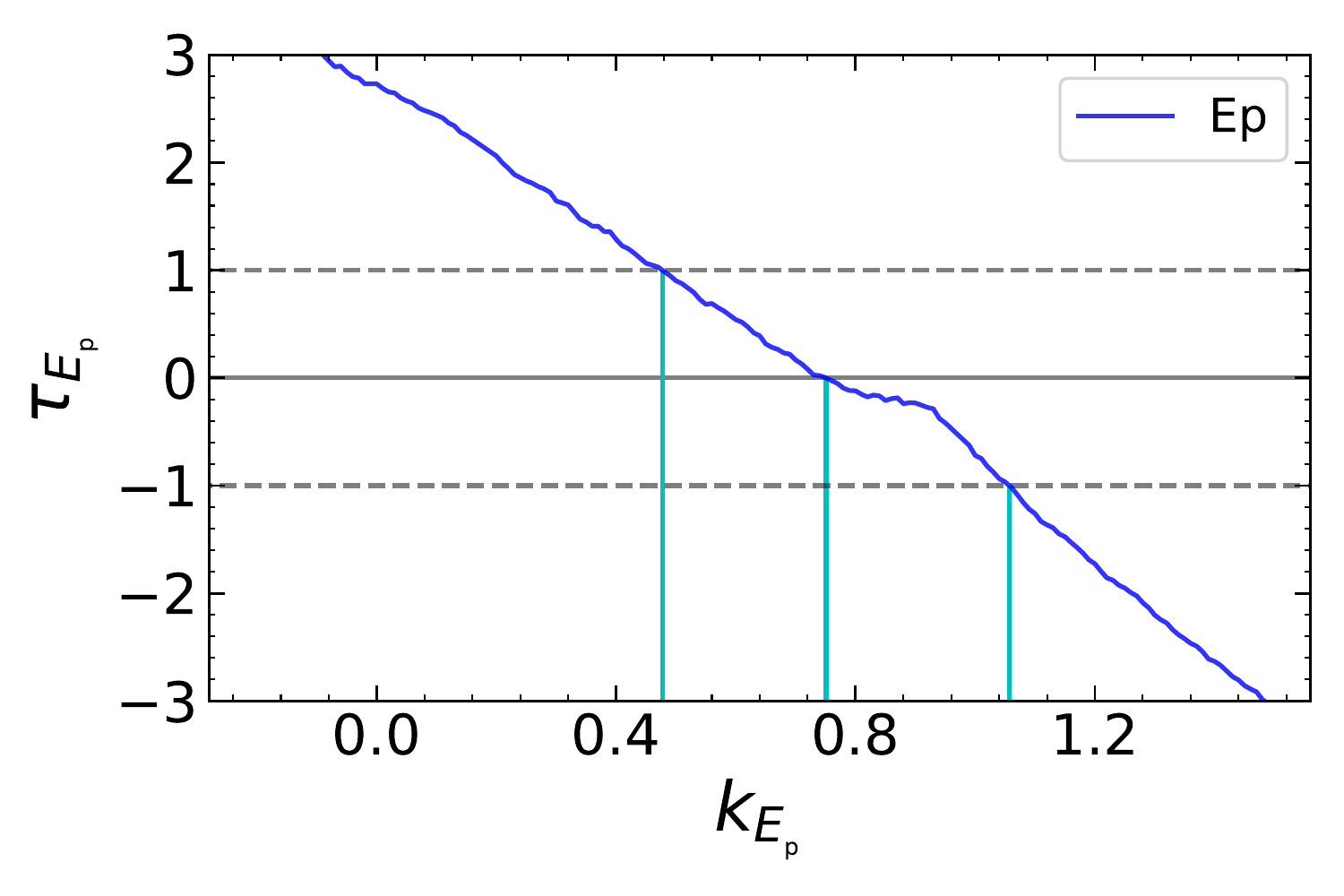}
	}
	
	\caption{ The relation between $\tau_{L_{X}}-k_{L_{X}}$ (left
panel), $\tau_{T_{a}}-k_{T_{a}}$ (middle panel), and $\tau_{E_{\rm{p}}}-k_{E_{\rm{p}}}$ (right
panel). The best $k$ value is determined by $\tau=0$, and the $1\sigma$ range of uncertainty is
derived from $|\tau|\le1$.}
	\label{fig:A2}
\end{figure}

\end{document}